\documentclass[a4paper,review]{elsarticle}

\usepackage[ruled,vlined]{algorithm2e}
\usepackage[breaklinks]{hyperref}
\usepackage{threeparttable}
\usepackage{mathtools}
\usepackage{microtype}
\usepackage{tabularx}
\usepackage{multirow}
\usepackage{booktabs}
\usepackage{csquotes}
\usepackage{amsmath}
\usepackage{amssymb}
\usepackage{lscape}
\usepackage[table]{xcolor}
\usepackage{array}
\usepackage{float}
\usepackage{todonotes}
\usepackage{hyperref}
\hypersetup{
colorlinks=true,
linkcolor=blue,
filecolor=blue,
urlcolor=blue,
citecolor=blue,
}

\usepackage{geometry}
\newcommand\extrafootertext[1]{%
    \bgroup
    \renewcommand\thefootnote{\fnsymbol{footnote}}%
    \renewcommand\thempfootnote{\fnsymbol{mpfootnote}}%
    \footnotetext[0]{#1}%
    \egroup
}
\makeatletter
\def\ps@pprintTitle{%
   \let\@oddhead\@empty
   \let\@evenhead\@empty
   \let\@oddfoot\@empty
   \let\@evenfoot\@oddfoot
}
\makeatother

\biboptions{numbers,sort&compress}
\begin{document}

\begin{frontmatter}

\title{An IoT-Based Framework for Remote Fall Monitoring}

\author[QU_EE]{Ayman Al-Kababji}\corref{correspondingauthor}
\cortext[correspondingauthor]{Corresponding author}
\ead{ayman.alkababji@ieee.org}

\author[DMU_AI]{Abbes Amira}
\ead{abbes.amira@dmu.ac.uk}

\author[QU_EE]{Faycal Bensaali}
\ead{f.bensaali@qu.edu.qa}

\author[QU_EE]{Abdulah Jarouf}
\ead{aj1506738@qu.edu.qa}

\author[QU_EE]{Lisan Shidqi}
\ead{ls1511236@qu.edu.qa}

\author[UO_ITEE]{Hamza Djelouat}
\ead{djelouat.hmz@gmail.com}

\address[QU_EE]{Department of Electrical Engineering, Qatar University, Doha, Qatar}
\address[DMU_AI]{Institute of Artificial Intelligence, De Montfort University, Leicester, United Kingdom}
\address[UO_ITEE]{Faculty of Information Technology and Electrical Engineering, University of Oulu, Oulu, Finland}

\begin{abstract}
Fall detection is a serious healthcare issue that needs to be solved. Falling without quick medical intervention would lower elderly’s chances of survival, especially if living alone. Hence, the need is there for developing fall detection algorithms with high accuracy. This paper presents a novel IoT-based system for fall detection that includes a sensing device transmitting data to a mobile application through a cloud-connected gateway device. Then, the focus is shifted to the algorithmic aspect where multiple features are extracted from 3-axis accelerometer data taken from existing datasets. The results emphasize on the significance of Continuous Wavelet Transform (CWT) as an influential feature for determining falls. CWT, Signal Energy (SE), Signal Magnitude Area (SMA), and Signal Vector Magnitude (SVM) features have shown promising classification results using K-Nearest Neighbors (KNN) and E-Nearest Neighbors (ENN). \textcolor{black}{For all performance metrics (accuracy, recall, precision, specificity, and F\textsubscript{1} Score), the achieved results are higher than 95\% for a dataset of small size, while more than 98.47\% score is achieved in the aforementioned criteria over the UniMiB-SHAR dataset by the same algorithms, where the classification time for a single test record is extremely efficient and is real-time\extrafootertext{}\extrafootertext{}\extrafootertext{}\extrafootertext{}\extrafootertext{}\extrafootertext{\textit{Published in Biomedical Signal Processing and Control}}}.
\end{abstract}


\end{frontmatter}


\section{Introduction} \label{sec1}
Enabling technology use in health sector brings many advantages that can save lives. The introduction of such systems for autonomous alarming and monitoring is significant in providing continuous supervision on patients’ and elderlies’ status. According to the United Nations (UN), people aged 65 and above constituted around 9.1\% of the total population in 2019 ($\sim$700 millions) \cite{Uni}. This percentage is expected to reach 15.9\% from the expected population (9.735 billions) in 2050 \cite{Uni}. Not only that, but for the first time in human history, in 2019, people aged 65 or more have exceeded the number of children below five \cite{Uni}. This is a life-changing outcome of high life expectancy around the world where previously mentioned systems present a solid solution for, particularly that there is an unavoidable shortage in healthcare personnel.

\textcolor{black}{Contrary to what has been mentioned earlier, healthcare systems have evolved significantly in the last century, where a lot of new technological advancements and clinical tests are implemented and discovered. However, because these systems have succeeded in raising people's life expectancy, another challenge arose as a result, which is the total increase in population, especially the elderly portion. From this point-of-view, healthcare systems have to re-adjust the provided services following a ``smart'' and automated methodology in order to accommodate the new needs that are currently arising.}

It is common between elderlies to be more physically vulnerable than other age groups, especially when it comes to falling incidents. One third of elderlies, over the age of 65, are reported to have fallen annually. Added to that, 50\% of elderlies who are aged above 80 also experience falls \cite{Com16}. These numbers are alarming since some of these incidents lead to bone fractures or, if severe enough, could end the elderly’s life. In addition, delayed medical intervention for elderlies who are immobile after a fall, especially those living alone, can further reduce their surviving chances and has a direct impact on their health outcomes \cite{Com16}. Thus, it is of the utmost importance to detect, and report falls of senior population in an attempt to reduce health complications due to the slow medical intervention.

\textcolor{black}{Reflecting on the current state of the world, in the COVID-19 pandemic era, healthcare personnel shortage is directly affecting, along with medical equipment shortage, the number of deaths due to the coronavirus. According to the Centers for Disease Control and Prevention (CDC), more than 80\% of the total deaths are for elderlies older than 65 in the US (up to 2\textsuperscript{nd} of January) \cite{COVID_19_CDC}. Furthermore, medical personnel are now more busy than ever with saving COVID-19 infected patients, making the continuous monitoring of elderly population a secondary issue at the moment, showing how important autonomous monitoring systems in these types of situations, where the elderly can be safely isolated at home, having continuous automated medical and professional monitoring. Looking at the COVID-19 pandemic from another aspect, according to Yamada et al. \cite{yamada2020effect}, because of the imposed lockdown, and other mobility restrictions that most countries are practicing to protect their citizens' welfare, these practices affect the physical activity levels of the elderly population, making them less active \cite{de2020falls}. Moreover, Goethals et al. \cite{goethals2020impact} argued earlier, before the distribution of various COVID-19 vaccines, that due to the absence of a vaccine that can provide protection to this population, many elderlies naturally declined participation in physical exercise in an attempt to avoid contracting the COVID-19 within the physical therapy centers \cite{de2020falls}, and they are right to fear for their lives, especially that elderlies are the most endangered age group. Nonetheless, the negative impact and damage from the reduced physical activity within this period will have its impact on elderlies, especially if involved in a rough event.}

\textcolor{black}{With the emergence of the Internet of Things (IoT) concept, wearable sensors/advanced smartphones, and powerful-enough attainable Central Processing Units (CPUs) and Graphical Processing Units (GPUs), automated monitoring systems are now achievable more than ever, and with an affordable price by the healthcare systems/elderlies. To illustrate the previous point, in an ideal scenario, a single smartphone can now serve as a sensing platform that acquires necessary sensory data as in \cite{UniMiB-SHAR}, a processing unit that classifies or predicts abnormal events as in \cite{cao2012falld}, and an alerting system that communicates with the caregiver in case of emergencies as in \cite{alkababji19}. This demonstrates that a complete automated monitoring system can be implemented on a device that the majority of, if not all, people have.}

\textcolor{black}{However, coming back to reality, the complexity of utilized algorithms, the continuous computations, and the classical battery limitations, render it to be impractical. Therefore, in an effort to build a complete feasible system, this paper introduces a novel fall detection system incorporating a sensing device, a gateway transceiver and a mobile application. Firstly, it investigates different systems mentioned in previous literature related to fall detection, focusing then on Wearable Sensing Device (WSD) systems showing that Continuous Wavelet Transform (CWT) is seldom used in such systems. Then, it shows the significance of extracting different features, emphasizing on CWT as a feature with high influence on fall detection. In addition, multiple classifiers, such as K-Nearest Neighbors (KNN), Extended-Nearest Neighbors (ENN), and Binary Decision Tree (BDT) are investigated. Finally, results on two different datasets are demonstrated, using only 3-axis accelerometer and in real-time, where a comparison is drawn with available products in the market in terms of their prices, and with ones found in literature in terms of achieved performance criteria.}

\textcolor{black}{Even though some of the investigated feature extraction algorithms and classifiers have been extensively utilized before in this domain, our main contributions can be summarized as follows:
\begin{itemize}
    \item Designing and concatenating CWT features, which are seldom utilized and optimized in literature, along with the Signal Energy (SE), Signal Vector Magnitude (SVM), and Signal Magnitude Area (SMA) to generate the best results. The exact values for the scale and the different wavelet functions for the CWT are identified, which help in generating the best results.
    \item Using the ENN classifier and with the specific optimal features' concatenation, especially that it is rarely utilized in the context of fall detection as the main classifier.
    \item Investigating optimal number of neighbors for both KNN and ENN, highlighting how ENN is less sensitive to variations within the number of neighbors exhibiting more robustness in its performance
    \item Finding the optimal features vector over a dataset of small size, which is then calculated over the UniMiB-SHAR dataset, showing the resilience of our hand-crafted engineered features and the extremely high achieved results +98.5\% for all performance criteria.
\end{itemize}}

The paper’s sections are organized as follows. Section \ref{sec2} shows fall detection-related work describing the concept and the sensors in use, and highlighting the pros and cons of each system. Section \ref{sec3} describes our proposed system for fall detection. Section \ref{sec4} demonstrates the methodology to examine different feature extraction algorithms with three different classifiers: KNN, ENN, and BDT. Section \ref{sec5} shows the results of different features combinations, different number of neighbors for KNN and ENN, and individual classifier’s performance. Moreover, a comparison is also drawn with available commercial products and other state-of-the-art studies. Finally, Section \ref{sec6} concludes the paper revisiting the most important highlights of this paper.

\section{Related Work}\label{sec2}
Looking into previous fall detection-related work in the last few years, many systems with a variety of monitoring ideas have been proposed and discussed. The main categories are based on the sensing device in use that include externally placed cameras, floor sensors, radars, WSD, smartphones, and more interestingly, through Wi-Fi Channel State Information (CSI).

\subsection{Cameras}
Camera-based systems use camera(s) and image processing techniques to detect elderlies' movements and analyze falls. It involves tracking the significant parts of the elderlies’ body. Omnidirectional-Camera as in \cite{Mia16} is used for fall detection purposes with input data as the RGB components of each frame. However, recently the usage of depth cameras seems to be the trend, where an Infrared (IR) emitter works together with a camera. Authors in \cite{Bia15,Sto15,panahi2018human,Abo18,mazurek2018use} utilize such cameras due to their ability in operating well under weak light conditions; since these cameras do not depend on the visible light spectrum. Hence, more reliable information can be attained. In \cite{Bia15}, the aim is to detect the 3D trajectory of the head joint, while in \cite{Sto15}, the goal is to monitor the individual’s vertical state in each image frame using Microsoft Kinect depth imaging sensor. Similarly, in \cite{panahi2018human}, Microsoft Kinect camera is utilized on 70 video samples, to detect falls using two different methods, Support Vector Machine-Machine Learning (SVM\_ML) and thresholding. However, in \cite{Abo18}, the authors claim that trying to locate the human key joints is difficult especially due to the occluded body parts and sudden posture changes. They proposed their own skeleton-free fall detection system by predicting individuals’ correct postures. Lastly, in \cite{mazurek2018use}, authors use an IR depth camera for detecting the elderly's silhouette, from which they extract two types of features, kinematic and mel-cepstrum features. Following, they feed the extracted features to multiple classifiers such as SVM\_ML, Artificial Neural Network (ANN) and Naïve Bayes classifiers for performance comparison.

Camera-based systems have a major concern; which is the privacy of the patient, since cameras are constantly recording private locations, hence, it might deter certain elderlies from opting into the system.

\subsection{Floor Sensors}
Floor sensors-based systems collect data using nearly unnoticeable sensors as in \cite{Zig09,Dah17,Rim10}. In \cite{Zig09}, the acoustic signals (i.e. vibration and sound), propagating through room’s floor, are acquired by accelerometer and microphone sensors stationed at the room’s corner. In a more recent publication, the authors in \cite{Dah17} use the INRIA-Nancy Smart Tiles prototype (around 100 tiles) where a 3-axis accelerometer is situated in the middle of the tile accompanied by four pressure sensors distributed at the corners. However, in \cite{Rim10}, a different type of input is acquired based on a conductive film under the floor powered to generate electric fields. A good analogy in explaining the mechanism behind this conductive film would be the way that touch screens work. The presence of human and resultant falls can be detected where a change in the film's impedance will be calculated \cite{Rim10}.

This type of systems, especially the one mentioned in \cite{Rim10}, is promising in nursing homes where elderlies have limited mobility. However, they need a pre-installation phase, before furnishing the rooms, where sensors should be equipped under the rooms’ floor.

\subsection{Radars}
Radar-based fall detection systems use radar sensors and require a base station for signals’ reception. Such systems, presented in \cite{Pet12,Gar15,SuB15,Jok18}, take advantage of the Doppler effect that signals exhibit when received/reflected by moving objects causing them to change in frequency. In \cite{Pet12} and \cite{Gar15}, the utilized feature generating highest accuracy is the Global Alignment (GA) Kernel belonging to the Least Squares Support Vector Machine (LS-SVM) classifier. Whereas in \cite{SuB15}, the extracted feature from the radar signal is the Discrete Wavelet Transform (DWT) and it is fed to KNN classifier with $K$ = 1. On the other hand, autoencoders (a form of ANN) are used in \cite{Jok18} to learn the features from the input radar signals, and the Softmax regression classifier to predict human’s movement. Softmax regression is the generalized case of logistic regression, and is used for multi-class classification scenarios.

\subsection{Wi-Fi CSI}
Wi-Fi CSI-based systems, as stated in \cite{Wan17}, aim to detect falls by observing indoor environment’s effects on the Wi-Fi’s radio signal through time delay, amplitude attenuation and phase shift. Thus, information estimation about the channel’s properties can be done. It utilizes the Wi-Fi devices as transmitters and have processing units as receivers such as computers. The CSI information is then extracted from the received packets \cite{Wan17}. Authors in \cite{Wan17,Wan171,Hua19} have all utilized this type of systems for detecting falls. Main difference between WiFall \cite{Wan17} and RT-Fall \cite{Wan171} is that the WiFall system depends only on the amplitude changes in the CSI, while RT-Fall investigates both amplitude and phase difference information produced by multiple antennas. In \cite{Hua19}, the proposed FallSense system uses Dynamic Template Matching (DTM) that relies initially on less training data that keeps updating during system’s usage. Moreover, the authors claim that their system is less complex since they use DTM with small training set instead of SVM and Hidden Markov Model (HMM) as in \cite{Wan17} and \cite{Wan171}.

It should be noted that Wi-Fi-based systems and radar-based systems use Radio Frequency (RF) similarly, however, Wi-Fi-based systems investigates human’s movements effects on CSI, while radar-based systems observe their effects on the signal’s frequency (Doppler effect). Since both systems relatively yield similar results, it is perhaps more convenient to use the Wi-Fi since it is pre-existing in homes whereas radar-based systems require extra installation.

Camera, floor, radar and Wi-Fi-based systems have two serious challenges in common. They would be useless if the elderly had a walk outside and fell. Meaning that for them to function correctly, the elderlies must stay indoor all the time, which imposes restricted mobility on elderlies since the sensors are static. In addition, since these systems rely on having a single individual or elderly indoor, pets and other people movement would generate false alarms \cite{Cip17}.

\subsection{WSD \& Smartphones}
The most common type of fall detection systems found in literature were the ones depending on WSD. It is safe to say that most researchers are opting to use the triaxial accelerometer as part of their system disregarding where the wearable sensor is placed as in \cite{Lee15,Pie15,Gib16,gibson2017matching,Azi17,Eju17,Wan172,deQ18,Saa19}. However, Ozcan et al. took a completely different approach where the device in use is a camera placed at the waist for entropy distance measurement \cite{Ozc16}. Comparison is not straightforward for fall detection systems of this type as there are many differences to account for. For instance, some researchers use only the 3-axis accelerometer sensor placed on the chest \cite{Gib16, gibson2017matching}, on the waist \cite{Azi17}, and on the thigh \cite{Saa19}. However, authors in \cite{Pie15} and \cite{deQ18} use a sensing device generating 3-axis accelerometer, gyroscope and magnetometer data, which was placed on the waist and the wrist, respectively. Hence, giving them more information about the orientation, velocity and displacement of the elderly. For authors in \cite{Eju17}, they utilize the 3-axis accelerometer and barometric pressure sensors, on the neck for sit-to-stand detection, while Wang et al. use the same type of sensor and placement for power-efficient signal features extraction \cite{Wan172}. In \cite{Eju17} Lee et al. talk about the usage of 3-axis accelerometer and gyroscope sensors on the elderly’s waist for detecting near-fall scenarios as well as falls \cite{Lee15}. 

More interestingly, since smartphones nowadays possess multiple sensors including 3-axis accelerometer and gyroscope, researchers utilized them for fall detection as both a sensing and a processing device. The idea formulated in early 2010s as smartphones processing capabilities started emerging at the time. A system that utilizes a smartphone is initially introduced in \cite{Wib13} where fall detection is implemented through a simple thresholding-based learning algorithm. In \cite{Kau15}, a more advanced learning algorithm such as Finite-State Machine (FSM) cascaded with SVM is used on the smartphone. Kau et al, interestingly argues that the power consumption resulting from running fall detection application is around 9\%, and it is the same consumption of a gaming application. Authors in \cite{Ker15}, use ANN on the smartphone for fall detection and, as expected, it drains a lot of power as shown in their results. FallDroid, developed by \cite{Sha19}, uses a thresholding-based algorithm followed by multiple kernel SVM learning algorithm. Implementing such systems, as in \cite{Kau15,Ker15,Sha19}, with high complexity on smartphones is mainly due to the technological enhancements mobile phones have undergone recently. 

Both WSD and smartphones fall detection systems are challenged by the limited amount of energy batteries can provide. Hence, it is not possible to deploy these systems on the same device 24/7. Moreover, both systems are susceptible to be forgotten by elderlies where forgetfulness is a common trait in that age category.

It is true that systems based on WSD can create uncomfortableness as well to the wearer, however, it is the most reliable source of data since they are coming from the elderlies themselves. In other words, existence of other people or pets near the elderly would not trigger a false positive as in the case of camera, floor, radar and Wi-Fi fall detection systems. Furthermore, they can be used to implement a system where the elderly does not have to be imprisoned in a house, which reduces mobility restrictions other systems impose. Table~\ref{tab:LR_summary} summarizes the reviewed technologies, highlighting their advantages and disadvantages.

\begin{table}[htbp]
    \caption{Literature review summary}
    \vspace{0.3cm}
    \label{tab:LR_summary}
    \centering
    \begin{tabular}{
m{0.15\linewidth}
>{\centering\arraybackslash}m{0.1\linewidth}
m{0.28\linewidth}
m{0.35\linewidth}
}
\toprule
\textbf{Technology} & \textbf{Studies} & \multicolumn{1}{c}{\textbf{Advantages}} & \multicolumn{1}{c}{\textbf{Disadvantages}} \\
\midrule
Cameras & \cite{Mia16,Bia15,panahi2018human,Sto15,Abo18,mazurek2018use} & Minimal human intervention & Privacy, multiple cameras \& indoor\\
Floor Sensors & \cite{Zig09,Dah17,Rim10} & Minimal human intervention \& more private than cameras & Frequent false alarms \cite{Abo18} \& indoor\\
Radars \& Wi-Fi CSI & \cite{Pet12,Gar15,SuB15,Jok18} \& \cite{Wan17,Wan171,Hua19} & More private than floor sensors \& cost-effective & High-sensitivity to noise \cite{SuB15} \& indoor\\
WSD \& Smartphones & \cite{Lee15,Pie15,Gib16,gibson2017matching,Azi17,Eju17,Wan172,deQ18,Saa19,Ozc16} & Immunity to noise \& minimal number of sensors, mobility & Depends on Elderlies' interaction, uncomfortable, needs recharging\\
\bottomrule
\end{tabular}
\end{table}

In this paper, extracting different features, from two pre-existing datasets \cite{Gib16,UniMiB-SHAR}, containing 3-axis accelerometer data is investigated. Furthermore, examining different feature combinations and focusing on CWT as a significant feature for fall detection where the WSD systems did not opt to use, except the one mentioned in \cite{Eju17}. Then, the optimal feature combination is fed to three different classifiers (KNN, ENN and BDT) and a majority Voting Machine (VM). KNN and ENN number of neighbors are varied where the optimal neighbor’s number is chosen based on specific performance criteria. Then, a comparison between individual classifiers is presented to show which one would be the best choice if only one is to be deployed. Finally, a comparison with commercial products and state-of-the-art studies is conducted, highlighting the competitiveness of the designed classifiers.

\section{Overall System} \label{sec3}
In this section, a description of the overall system, from both its hardware and software aspects, is presented elaborating on the role of each device.

\subsection{Hardware Implementation}
\begin{figure}[!ht]
    \centering
    \includegraphics[trim={0in 0in 0in 0in},clip,width=0.7\linewidth]{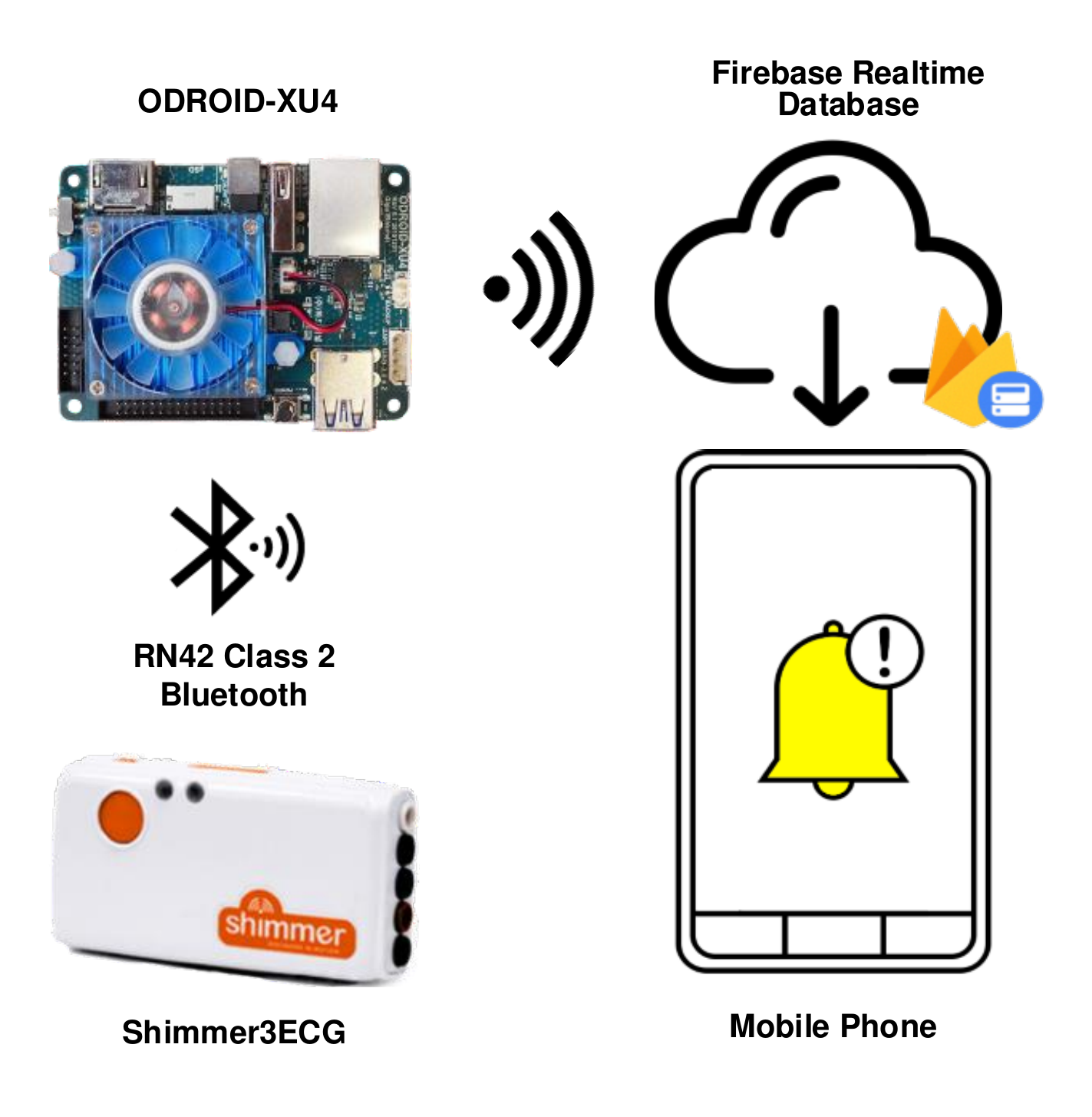} 
    \caption{Overall system design}
    \label{fig:system_design}
\end{figure}

As illustrated in Figure~\ref{fig:system_design}, several components and communication protocols are involved. A WSD (Shimmer3ECG) transmits acquired 3-axis accelerometer and electrocardiogram (ECG) data to a gateway device (ODROID-XU4) through Bluetooth. 

\begin{figure}[ht!]
    \centering
    \includegraphics[trim={0in 0.3in 0in 0in},clip,width=0.75\linewidth]{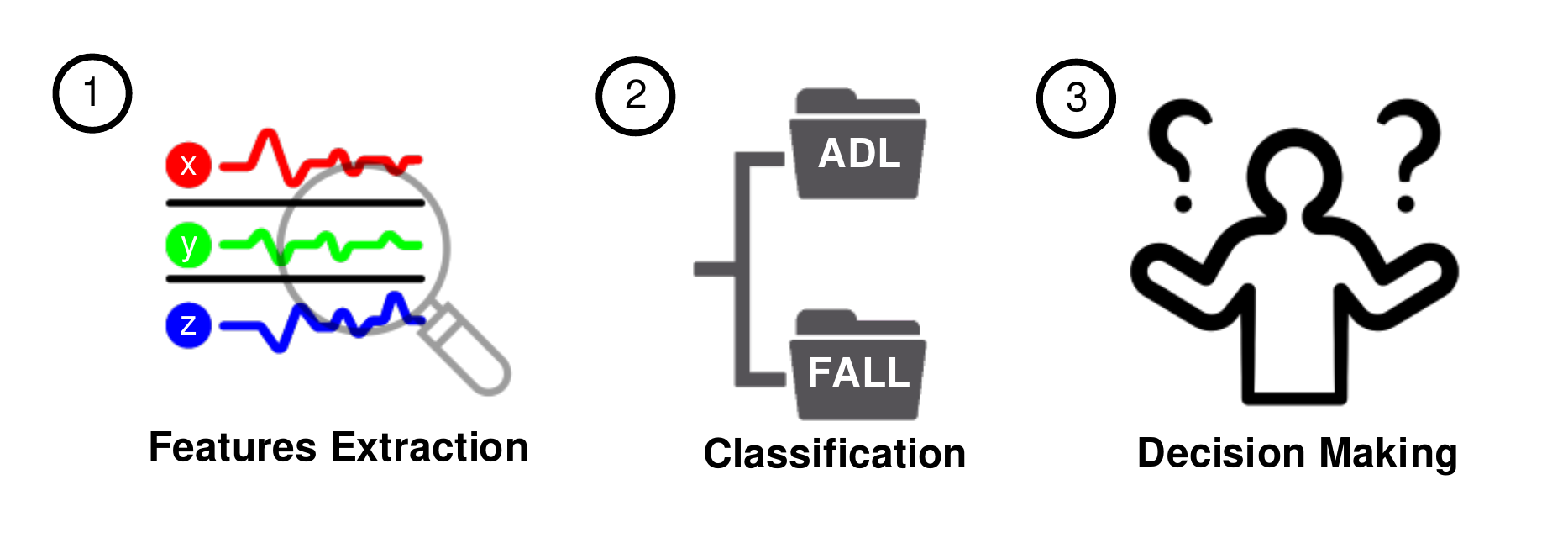} 
    \caption{Received 3-axis accelerometer signal processing steps}
    \label{fig:algorithm_steps}
\end{figure}

On ODROID-XU4, as shown in Figure~\ref{fig:algorithm_steps}, feature extraction algorithms are applied to the received data where the most significant features are looked for. Then, they are fed to chosen classifiers where they are trained to distinguish between Activities of Daily Living (ADL) and fall events. If a fall is detected, ODROID-XU4 will send an alarm to caregivers’ smartphones through a well-structured cloud database channel. Hence, ensuring a quick response from caregivers to aid the elderly in danger.

The Shimmer3ECG sensor is used mainly due to its small size, its power efficiency, programmability, and its capability in streaming ECG and accelerometer data in real-time simultaneously \cite{Gib16}. On the other hand, having an octa-core Exynos5422 big.LITTLE processor, a 2GB LPDDR3 RAM, and an advanced Mali GPU are the main reasons behind choosing ODROID-XU4 \cite{ODR}. In addition, a Wi-Fi dongle can be connected to the ODROID-XU4 to transmit data to the cloud.

\subsection{Software Implementation}
\begin{figure}[H]
    \centering
    \includegraphics[trim={0.5in 0in 0.2in 0in},clip,width=0.9\linewidth]{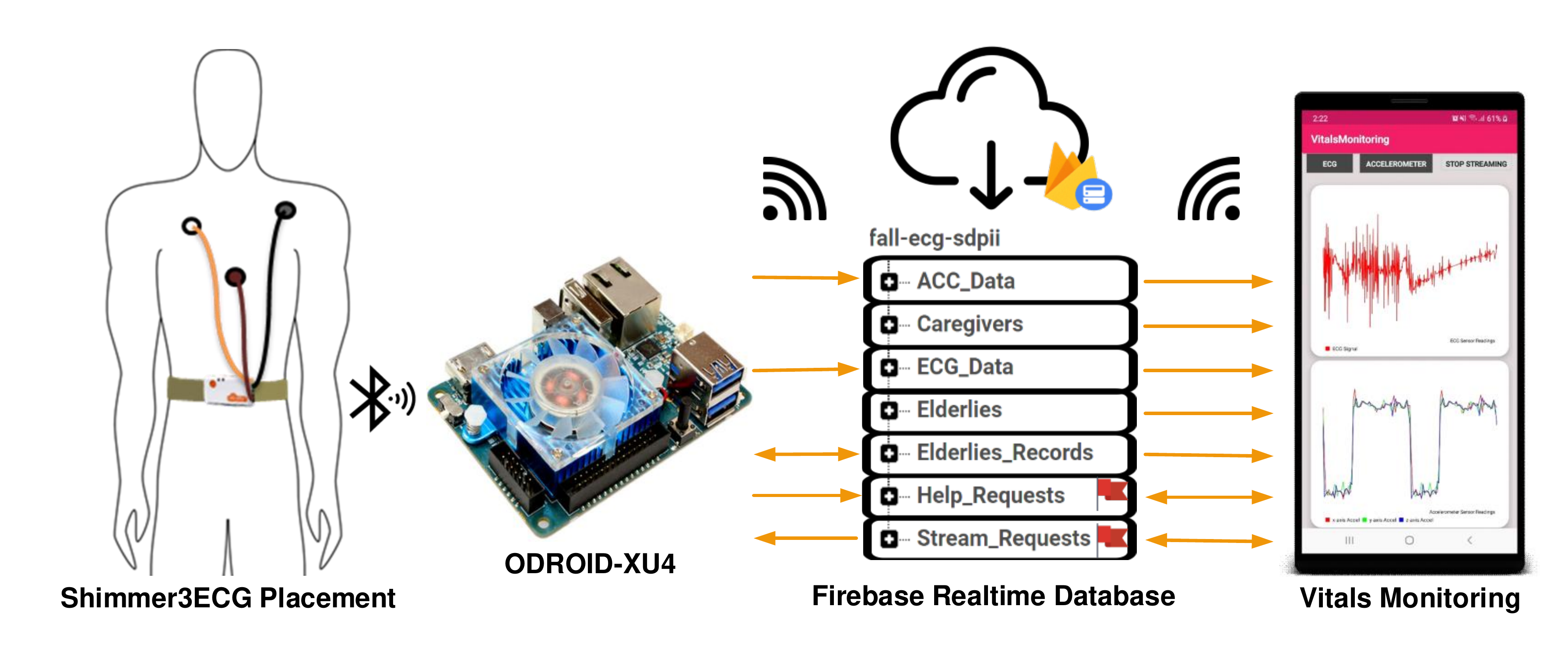} 
    \caption{Communication system}
    \label{fig:communication_setup}
\end{figure}

The proposed communication structure in Figure~\ref{fig:communication_setup} highlights interlinked entities within the cloud-hosted database platform, and illustrates the role of both the ODROID-XU4 and the user’s mobile application for data visualization \cite{alkababji19}. As illustrated by Figure~\ref{fig:communication_setup}, the  elderly-to-caregiver communication system involves the link from the Shimmer3ECG, the ODROID-XU4 as the gateway and the cloud-hosted Firebase Realtime database up to the Vitals Monitoring mobile application.

The Firebase platform provides a back-end service that allows for the real-time storage of large datasets within a database that synchronizes all clients with internet access.

\begin{figure}[ht!]
    \centering
    \begin{tabular}{c}
        \includegraphics[trim={0in 0in 0in 0in},clip,width=0.6\linewidth]{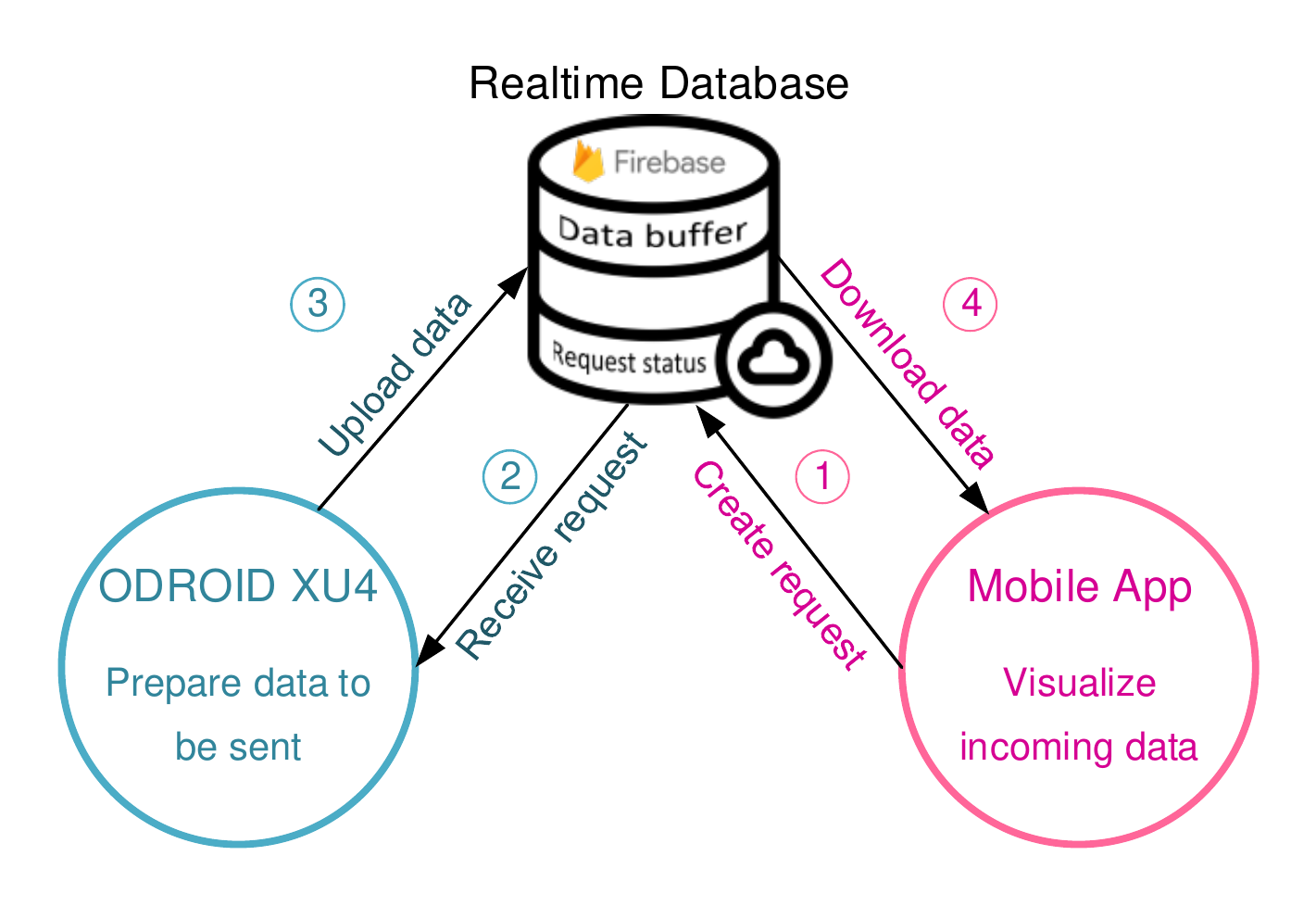}\\
        (a)\\
        \includegraphics[trim={0in 0in 0in 0in},clip,width=0.6\linewidth]{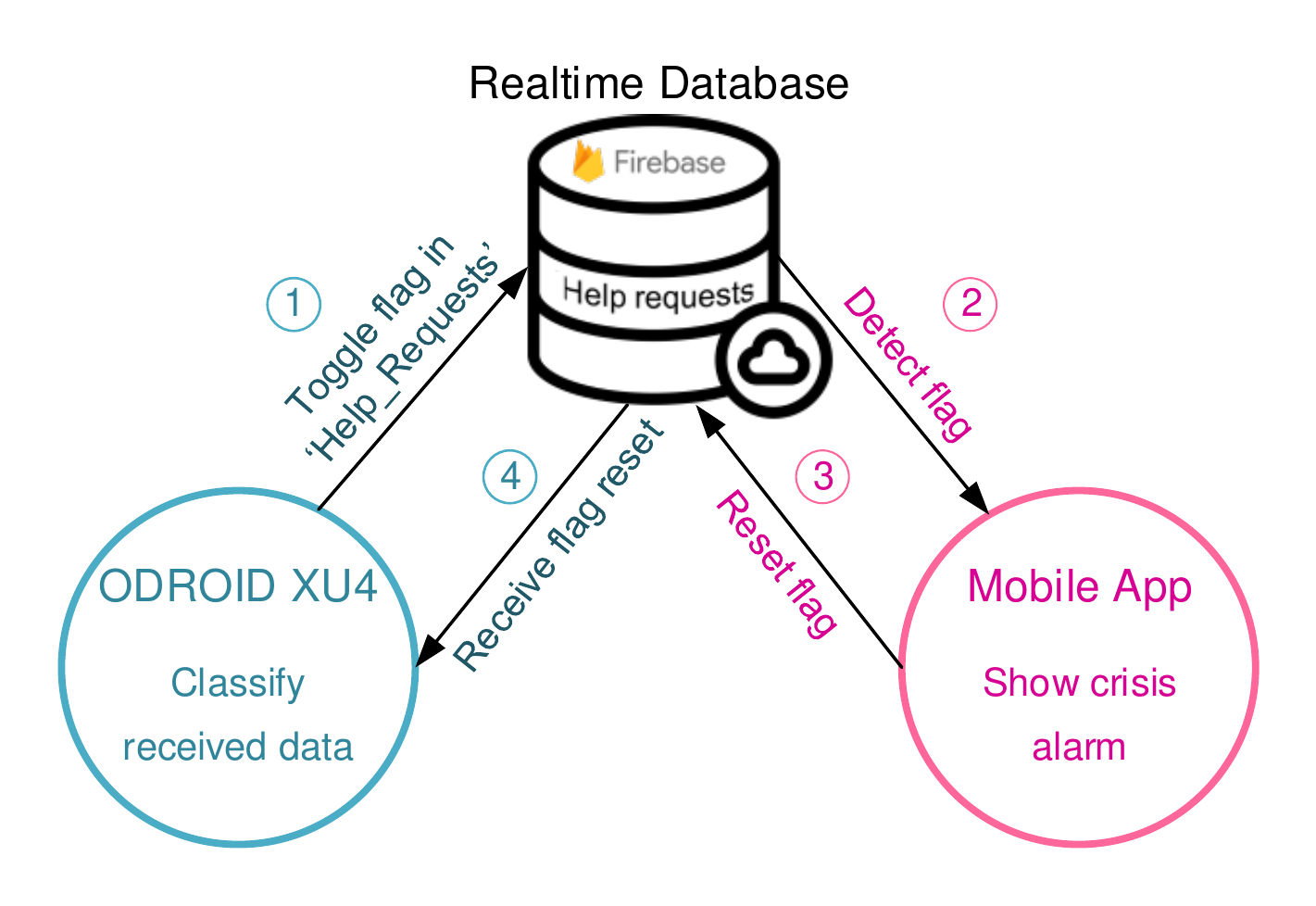}\\
        (b)
    \end{tabular}
    \caption{Communication protocols: a) data streaming request; b) alerting caregivers request}
    \label{fig:communication_protocols}
\end{figure}

The database is meant to act as a temporary buffer for the uploaded signals data, which should be accessed remotely by another client device with authenticated access, in this case, a mobile application that displays the stream of accelerometer and/or ECG data. The uploaded data are updated continuously with new values during streaming, while the mobile application ``listens'' to data changes on the cloud and handles them appropriately. The interactions of the different entities with the cloud-hosted database are illustrated in Figure~\ref{fig:communication_protocols} (a) for data visualization, and Figure~\ref{fig:communication_protocols} (b) for alerting caregivers.

A separate Python program would be running by the ODROID-XU4 acting as a client device, standing by for when data streaming is required. To communicate with the cloud database, an appropriate helper library (Pyrebase) in Python is utilized where it provides necessary Application Programming Interface (API) to communicate with the Firebase cloud. The Python program would receive the data pool to be uploaded and stores them temporarily. Using the Pyrebase library, it establishes the proper channel connection with the cloud database through internet, where the program then sends the data. As a result, synchronization happens when the data are put into the cloud database. 

\begin{figure}[hb!]
    \centering
    \includegraphics[trim={0in 0in 0in 0in},clip,width=0.8\linewidth]{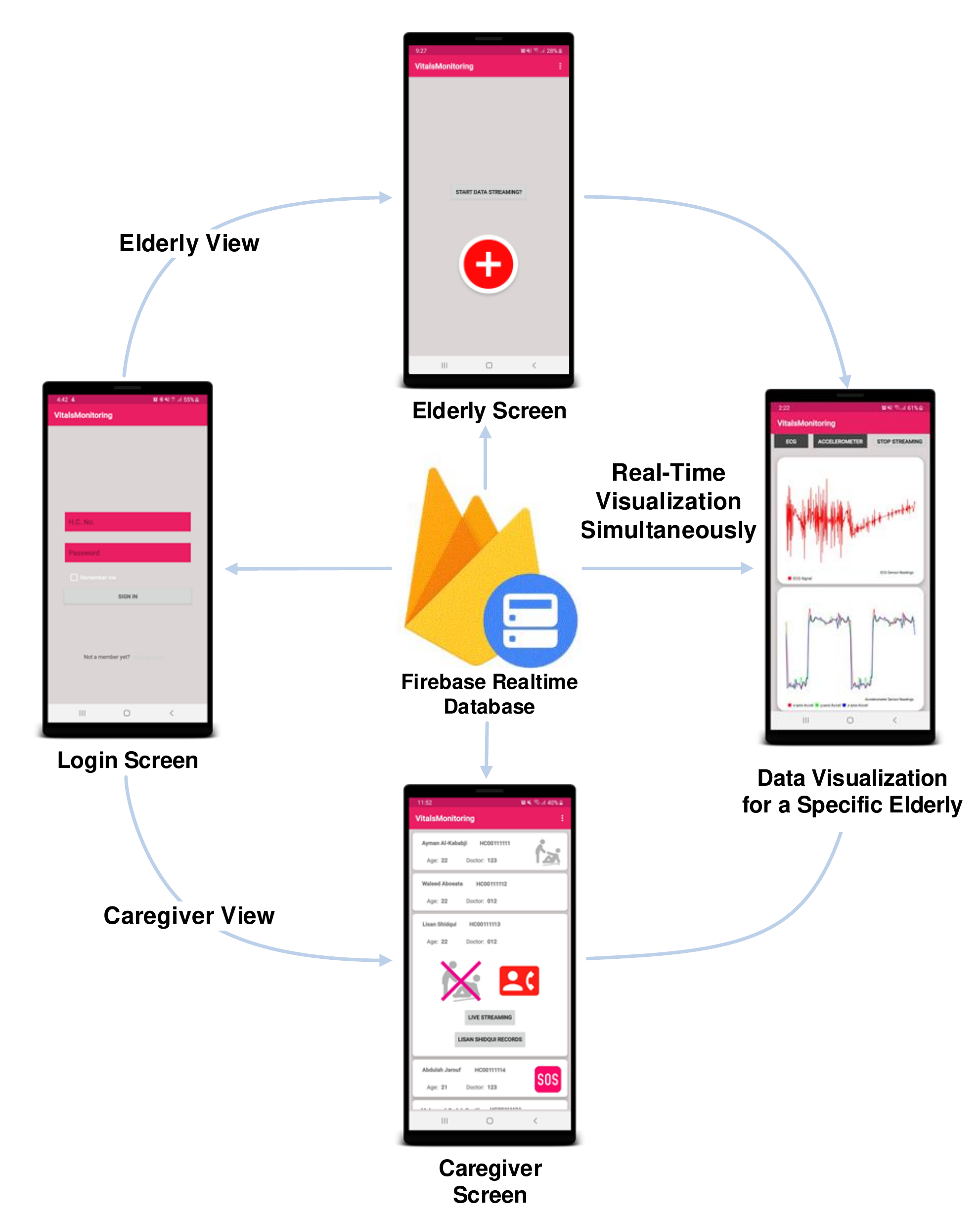} 
    \caption{Application usage cycle}
    \label{fig:usage_cycle}
\end{figure}

The developed mobile application (Vitals Monitoring) is intended to run on Android devices as a proof of concept. Its role revolves mainly on providing data visualization for both elderlies and caregivers and facilitating an alarming feature for caregivers in case of an emergency from any elderly. There are two separate portals of the Vitals Monitoring application, categorized as the Elderly View and the Caregiver View as shown in Figure~\ref{fig:usage_cycle}. 

Caregiver View provides a list of all registered elderlies, and caregivers can see any of the elderlies' details. Moreover, they can receive and visualize the streamed data from the elderly’s side, or even call the elderly for a quick check-up. In the Elderly View, he/she can visualize the data that is streamed from the Shimmer3ECG connected to him/her, as well as ask for help from the caregiver in time of need.

\begin{figure}[ht!]
    \centering
    \includegraphics[trim={0in 0.5in 0in 0in},clip,width=1\linewidth]{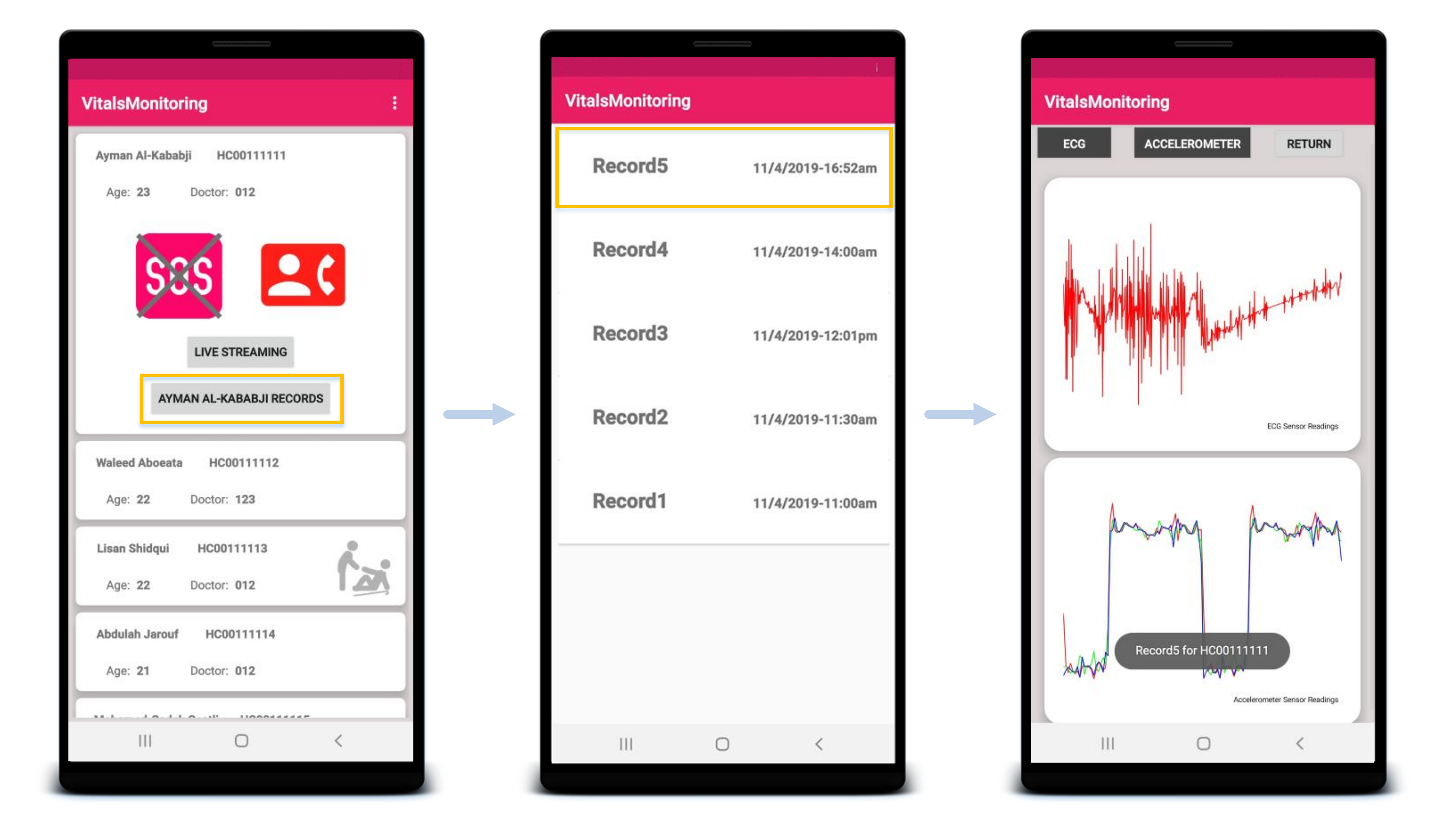} 
    \caption{Previous falling records}
    \label{fig:falling_records}
\end{figure}

Lastly, all users’ data are available on the cloud database for user authentication. Furthermore, previous fall records for each elderly, containing ECG and 3-axis accelerometer data stored at falling time, are saved under the Elderlies\textunderscore Records path, illustrated in Figure~\ref{fig:communication_setup}. As shown in Figure~\ref{fig:falling_records}, only caregivers have access to these records for further analysis.

\section{Methodology} \label{sec4}
In this section, the pursued methodology to reach the presented results in the following section are discussed. Firstly, the used dataset is tabulated showing the total number of fall and ADL records. Secondly, the features that are extracted from the previously mentioned dataset are discussed. Thirdly, the machine learning classifiers in use are explained and the criteria for classifiers’ performance check are also demonstrated. These criteria are important as they assist in validating the classifier prediction performance.

\subsection{Utilized Training Datasets}
Supervised machine learning algorithms depend heavily on having a well-structured and properly labeled dataset. Building such dataset requires significant amount of time, thus, a part of the dataset generated by authors in \cite{Gib16} is used as it employed the previous version of the used WSD in this paper. The content used from the dataset is described in Table~\ref{tab:dataset_content_minor}. \textcolor{black}{Moreover, after the verification of optimal features and classification algorithms over the Gibson et al. dataset \cite{Gib16}, they are trained and tested on a more comprehensive dataset named ``University of Milano Bicocca Smartphone-based Human Activity Recognition'' (UniMiB-SHAR) \cite{UniMiB-SHAR}, where it has a total of 11,771 records (ADL: 7,579 and Falls: 4,192) collected from a Samsung Galaxy Nexus I9250 equipped with a Bosh BMA220 acceleration sensor. Further details about that dataset and its distribution are presented in Table~\ref{tab:dataset_content_major}.}

\begin{table}[!ht]
    \caption{Gibson et al. Dataset content \cite{Gib16}}
    \vspace{0.3cm}
    \label{tab:dataset_content_minor}
    \centering
    \begin{tabular}{>{\centering\arraybackslash}m{0.12\linewidth} p{0.5\linewidth} >{\centering\arraybackslash}p{0.2\linewidth}}
\toprule
\multicolumn{1}{c}{\textbf{Type}} & \multicolumn{1}{c}{\textbf{Event}} & \multicolumn{1}{c}{\textbf{Records}} \\
\midrule

\multirow{11}{*}{\textbf{ADL}}
& Jumping (3 times) & 6 \\
& Jumping (1 time) & 6 \\
& Lie down from sitting position & 6 \\
& Lie down from sitting position quickly & 6 \\
& Running & 15 \\
& Sitting on chair & 7 \\
& Sitting on chair quickly & 6 \\
& Standing up & 6 \\
& Standing up quickly & 6 \\
& Walking & 22 \\
& Walking quickly & 6 \\
\midrule
\multicolumn{2}{c}{\textbf{Total ADL Events}} & \textbf{92} \\
\midrule
\multirow{8}{*}{\textbf{Fall}}
& Soft front fall & 19 \\
& Soft back fall & 19 \\
& Soft left fall & 19 \\
& Soft right fall & 19 \\
& Strong front fall & 15 \\
& Strong back fall & 15 \\
& Strong left fall & 15 \\
& Strong right fall & 15 \\
\midrule
\multicolumn{2}{c}{\textbf{Total Fall Events}} & \textbf{136} \\
\midrule
\multicolumn{3}{c}{\textbf{Dataset size = 228}} \\
\bottomrule
\end{tabular}
\end{table}

\begin{table}[!ht]
    \caption{\textcolor{black}{UniMiB-SHAR Dataset content \cite{UniMiB-SHAR}}}
    \vspace{0.3cm}
    \label{tab:dataset_content_major}
    \centering
    \begin{tabular}{>{\centering\arraybackslash}m{0.12\linewidth} p{0.5\linewidth} >{\centering\arraybackslash}p{0.2\linewidth}}
\toprule
\multicolumn{1}{c}{\textbf{Type}} & \multicolumn{1}{c}{\textbf{Event}} & \multicolumn{1}{c}{\textbf{Records}} \\
\midrule

\multirow{11}{*}{\textbf{ADL}}
& Standing up from sitting & 153 \\
& Standing up from lying & 216 \\
& Walking & 1,738 \\
& Running & 1,985 \\
& Going upstairs & 921 \\
& Jumping & 746 \\
& Going downstairs & 1,324 \\
& Lying down from standing & 296 \\
& Sitting down & 200 \\
\midrule
\multicolumn{2}{c}{\textbf{Total ADL Events}} & \textbf{7,579} \\
\midrule
\multirow{8}{*}{\textbf{Fall}}
& Falling forward & 529 \\
& Falling right & 511 \\
& Falling backward & 526 \\
& Falling left & 534 \\
& Hitting obstacle & 661 \\
& Falling with protection strategies & 484 \\
& Falling backward sitting on chair & 434 \\
& Syncope & 513 \\

\midrule
\multicolumn{2}{c}{\textbf{Total Fall Events}} & \textbf{4,192} \\
\midrule
\multicolumn{3}{c}{\textbf{Dataset size = 11,771}} \\
\bottomrule
\end{tabular}
\end{table}

Both datasets contain ADL of different types covering some of the most common daily activities elderlies do. For the fall events, they recorded falls in all directions and with different scenarios to maximize the possibility of detecting these events regardless of their cause or direction. Sampling frequency for both datasets was set to be 50 Hz and the period of each measured record (event) is 2 seconds for the Gibson et al. \cite{Gib16} and 3 seconds for the UniMiB-SHAR dataset \cite{UniMiB-SHAR}.

For our case, we consider the two classes classification issue, a fall or a non-fall (ADL) event. All soft and strong fall events were grouped in a single category under “fall” and the same is applied to ADL for the Gibson et al. dataset. Similar approach is applied over the UniMiB-SHAR dataset, where the classification of fall and non-fall events is named ``AF-2'' in their work \cite{UniMiB-SHAR}.

\subsection{Feature Extraction Algorithms}
Depending on the problem in hand and the signal in measurement, features will significantly vary in what they represent and the technique to extract them.

Triaxial accelerometer data are to be dealt with and fall events are to be detected here. Hence, the following features, existing in time, frequency, and time-frequency domains are chosen as it is foreseen that they would contribute the most.

\subsubsection{CWT}
CWT is an excellent decomposition tool where a non-static wave’s changing properties will be captured in small wavelets localized in time \cite{Con18}. These wavelets are shifted and scaled versions of the original mother wavelet. CWT outdoes the famous Short-Time Fourier Transform (STFT) in providing varying time windows for different frequencies. By allocating smaller time windows for high frequencies and large time windows for lower frequencies, this increases the resolution in the time-frequency domain as frequencies become higher \cite{Rio91}. Equation \eqref{eq:CWT} shows the general equation for CWT, which gives features in the time-frequency domain:

\begin{equation} \label{eq:CWT}
    {CWT}_x\left(a,b\right)=\frac{1}{\sqrt{\left|a\right|}}\int_{-\infty\ }^{\infty}{x\left(t\right)\psi\left(\frac{t-b}{a}\right)dt\ }
\end{equation}
\begin{flalign*}
\text{where:}\quad
&\text{$a$ is the scaling factor}& \\
&\text{$b$ is the translational factor}&\\
&\text{$x(t)$ is the processed signal}&\\
&\text{$\psi(t)$ is a mother wavelet function (filter)}&
\end{flalign*}

Triaxial accelerometer data representing events (whether ADL or fall events) have unique distinguishable traits from one another. For instance, the signal of a person jumping three times would have a repetitive behavior, while if he/she was to fall, the signal would have a high peak and a semi-constant signal afterwards (for one second at least). Hence, using CWT, as a feature to extract these traits would enhance the performance of the used classifiers.

\subsubsection{Signal Vector Magnitude (SVM)}
SVM is a time domain feature that calculates the total acceleration magnitude generated by the existing triaxial acceleration data \cite{Abd18}. It shows how significant the change in acceleration in a moment of time, disregarding on which axis that change was, through measuring the magnitude using equation \eqref{eq:SVM_i}:
\begin{equation} \label{eq:SVM_i}
    SVM_i=\sqrt{{Ax}_i^2+{Ay}_i^2+{Az}_i^2}
\end{equation}
\begin{flalign*}
\text{where:}\quad
&\text{$SVM_i$ is the $i^{th}$ acceleration vector magnitude}& \\
&\text{$Ax_i$, $Ay_i$ \& $Az_i$ are the $i^{th}$ acceleration element from $x$, $y$ \& $z$}& \\ &\text{accelerometer vectors respectively}&
\end{flalign*}

It is worth mentioning that the existence of the time parameter, through the order of samples, plays an important role in classifying an event whether it is a fall or not.

\subsubsection{Total \texorpdfstring{$\mid$SVM$\mid$}{|SVM|}}
Total $\mid$SVM$\mid$ feature is based on the previously mentioned SVM feature. The difference appears in summing the absolute value of all calculated SVMs as shown in equation \eqref{eq:Total_SVM}:

\begin{equation} \label{eq:Total_SVM}
    Total\ |SVM|=\sum_{i=1}^{M}|{SVM}_i|
\end{equation}
\begin{flalign*}
&\text{where: $M$ is the number of samples (the discrete equivalent of time interval)}&
\end{flalign*}

By doing so, this feature becomes independent from time, since all the \(\lvert SVM_i\rvert\)s are summed into one single feature.

\subsubsection{Signal Magnitude Area (SMA)}
SMA is another feature used to capture the observed amount of change from the acquired triaxial accelerometer data. Equation \eqref{eq:SMA} calculates this feature \cite{Sha19}:

\begin{equation}\label{eq:SMA}
    SMA=\sum_{i=1}^{M}\big(|Ax_i|+|Ay_i|+|Az_i|\big)
\end{equation}

\subsubsection{Triaxial Accelerometer Data Range}
The range of each accelerometer axis in time domain is the difference between the maximum and minimum values found in the tested record. They are calculated through equation \eqref{eq:range}:
\begin{equation}\label{eq:range}
\begin{split}
    Range_x=max(Ax)-min(Ax)\\
    Range_y=max(Ay)-min(Ay)\\
    Range_z=max(Az)-min(Az)
\end{split}
\end{equation}
\begin{flalign*}
&\text{where: $Ax$, $Ay$ \& $Az$ are the $x$, $y$ \& $z$ accelerometer vectors respectively}&
\end{flalign*}

\begin{figure}[ht!]
    \centering
    \includegraphics[trim={0.3in 0in 0.3in 0in},clip,width=0.8\linewidth]{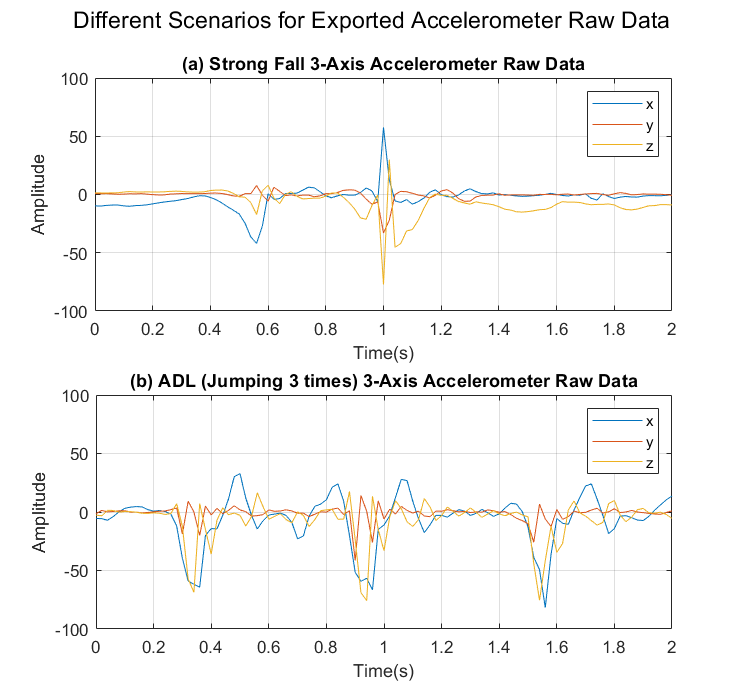} 
    \caption{(a) Strong fall event vs. (b) ADL (Jumping three times) event}
    \label{fig:fall_vs_adl}
\end{figure}

This feature holds information of how big the difference in each axis, but it does not necessarily mean that the record is a fall event if the range is large (see Figure~\ref{fig:fall_vs_adl}).

\subsubsection{Signal Energy (SE)}
SE is a frequency domain feature that is calculated using equation \eqref{eq:SE}:

\begin{equation} \label{eq:SE}
    E_x=\sum_{i=1}^{N}a_{x,i}^2,\ E_y=\sum_{i=1}^{N}a_{y,i}^2,\ E_z=\sum_{i=1}^{N}a_{z,i}^2
\end{equation}
\begin{flalign*}
\text{where:}\quad
&\text{\textit{a\textsubscript{x,i}}, \textit{a\textsubscript{y,i}} \& \textit{a\textsubscript{z,i}} are the \textit{i\textsuperscript{th}} Fast Fourier Transform (FFT) coefficients of the \textit{x}, \textit{y} \& \textit{z} axes, respectively}&\\
&\text{\textit{E\textsubscript{x}}, \textit{E\textsubscript{y}} \& \textit{E\textsubscript{z}} are the energy features for the \textit{x}, \textit{y} \& \textit{z} axes, respectively}&\\
&\text{$N$ is the total number of FFT coefficients per axis}
\end{flalign*}

The frequency coefficients are calculated for the 3-axis accelerometer data using FFT. Then, the resulting coefficients in each axis will be squared and summed to find the energy exerted by each axis \cite{Erd16}.

Table~\ref{tab:feature_extraction_summary} summarizes the extracted features in MATLAB and in which domain each feature resides. \textcolor{black}{It is worth noting that the aforementioned feature extraction algorithms are used in the literature, some more than the others, such as the SVM where it is almost used in every fall detection system that utilizes acceleration data. However, to the best of our knowledge, CWT features are not investigated deeply in the manner we portray, highlighting the effect of varying parameters such as the scale and the utilized wavelet function. Moreover, an investigation is carried out to find the best concatenation of such features, evaluating their overall performance and highlighting the best combination that yields the most promising results.}

\begin{table}[htbp]
    \caption{Feature extraction algorithms summary}
    \vspace{0.3cm}
    \label{tab:feature_extraction_summary}
    \centering
    \begin{tabular}{
m{0.2\linewidth}
m{0.47\linewidth}
>{\centering\arraybackslash}m{0.22\linewidth}}
\toprule
\textbf{Extraction Method} & \multicolumn{1}{c}{\textbf{Brief Explanation}} & \multicolumn{1}{c}{\textbf{Domain}} \\
\midrule
\textbf{CWT} & Finds the CWT equivalent for a given signal & Time-Frequency \\
\textbf{SVM} & Resultant magnitude vector of the 3-axial data & Time \\
\textbf{Total $\mid$SVM$\mid$} & Sum all absolute values of SVMs & Time \\
\textbf{SMA} & Area under the acceleration curve & Time \\
\textbf{Accelerometer Data Range} & Subtract the minimum value from the maximum value per axis & Time \\
\textbf{SE} & Calculate FFT and sum the square of each axis data separately & Frequency \\

\bottomrule
\end{tabular}
\end{table}

\subsection{Machine Learning Classification Algorithms}
The extracted features, as shown in Table~\ref{tab:feature_extraction_summary}, are fed to the following implemented supervised machine learning classifiers: KNN, ENN, and BDT. Then, the outputs of each classifier are inserted into VM that yields an output based on the majority of the aforementioned classifiers. 

\subsubsection{KNN}
KNN algorithm is an instance-based classification algorithm that uses the whole dataset to classify events based on the provided features. It calculates the distance between features of the new record with the existing training dataset features. After that, it uses the $K$ nearest neighbors in predicting which class does the new record belong to.

\subsubsection{ENN}
ENN has the same basis of operation as KNN; however, it differs from KNN in that it has a two way of communication. Meaning that it checks the nearest neighbors to the test record, and at the same time, takes into consideration the neighbors that see the test record as one of their nearest neighbors \cite{Tan15}. \textcolor{black}{Because of that, the algorithm builds weighted KNN map in a pre-processing ``training'' step that is done only once and is then utilized in the testing phase.}

\subsubsection{BDT}
BDT is a model-based classification algorithm. It builds a tree structure where in each node a decision must be made between two choices (binary). It breaks the dataset down into smaller subsets, at the same time, the associated decision tree keeps developing until it reaches a leaf (decision).

\subsubsection{VM}
VM checks what each classifier yielded, and based on that, makes the decision to detect an ADL or a fall event. Thus, it is not a classifier by itself.

Both KNN and ENN are convenient to be used here as the datasets are relatively small in size. For problems with much bigger datasets, it is not recommended to use such algorithms. They calculate the distance between the new incoming record’s features and the features of all the records, making them computationally expensive for huge datasets.

To test whether these classifiers are accomplishing satisfying results, calculating the following quantities, shown by their equations (\ref{eq:accuracy}-\ref{eq:specificity}) is a must:
\begin{equation}\label{eq:accuracy}
    Accuracy\ (Ac)=\dfrac{TP+TN}{TP+TN+FP+FN}\ast100\%
\end{equation}
\begin{equation}\label{eq:recall}
    Recall\ (Re)=\dfrac{TP}{TP+FN}\ast100\%
\end{equation}
\begin{equation}\label{eq:precision}
    Precision\ (Pr)=\dfrac{TP}{TP+FP}\ast100\%
\end{equation}
\begin{equation}\label{eq:F1_score}
    F_1\ Score\ (F_1)=2\ast\dfrac{Precision\ast Recall}{Precision+Recall}
\end{equation}
\begin{equation}\label{eq:specificity}
    Specificity\ (Sp)=\dfrac{TN}{TN+FP}\ast100\%
\end{equation}
\begin{flalign*}
\text{where:}\quad
&\text{True Positive ($TP$): A fall occurs, and the system properly detects it}&\\
&\text{False Positive ($FP$): The system detects a fall although it did not occur}&\\
&\text{True Negative ($TN$): An ADL is performed, and the system does not detect a fall}&\\
&\text{False Negative ($FN$): A fall occurs but the system does not detect it}&
\end{flalign*}

Accuracy measures how good the classifier’s predictions are with respect to the whole test set. It would be convenient if the dataset was structured to have classes of equal number of records. For instance, collecting 200 falling events and 200 ADL events would be an ideal case where accuracy could indicate if the classifier is performing well. Calculating other performance criteria such as recall and precision will surely help in assessing the classifier’s performance, especially if an equal number of records for each class is not available.

Recall measures the percentage of falls that were “correctly” detected from the fall test set. Precision on the other hand weighs how many of the detected falls were actual true falls. Moreover, F\textsubscript{1} Score is also important especially if recall and precision were not showing promising results and a choice is to be made between different models/classifiers. However, if both were high enough, F\textsubscript{1} Score will also have high percentage. Lastly, specificity shows how many ADL are “correctly” predicted from the overall ADL test set. It is similar to recall but for ADL events instead of fall events.
The pseudo code describing the methodology of generating the results in the following section is outlined by Algorithm \ref{algo1}.
\newline

\begin{algorithm}[H] \label{algo1}
\SetAlgoLined
\KwResult{Accuracy, Recall, Precision, F\textsubscript{1} Score \& Specificity}
folds := 5\;
Import dataset\;
Extract features (CWT, SVM, Total $\mid$SVM$\mid$, SMA, range, SE)\;
\While{$K\leq$ 17 \& $E\leq$ 17}{
    Randomize dataset\;
     \While{$i \leq folds$}{
        1. Divide dataset to 70\% training and 30\% testing\;
        2. Train classifiers (KNN, ENN, BDT)\;
        3. Predict testing data labels \& evaluate performance metrics\;
  }
}
Calculate the average of performance criteria\;
\caption{Classification pseudo code}
\end{algorithm}

\section{Results \& Discussion of Different Feature Extraction Combinations \& Different Classifiers}\label{sec5}

\subsection{\textcolor{black}{Results Analysis}}

The results are based on randomizing the dataset and splitting it into 70\% for training and 30\% for testing \cite{Ist16} by convention when the number of records in the dataset is low. Furthermore, the data are folded five times and for each iteration, accuracy, specificity, sensitivity, recall, and F\textsubscript{1} Score are calculated. The results shown in this paper are the average of these five folds.

The number of extracted features from the triaxial accelerometer data for both datasets from \cite{Gib16} (2 seconds records) and UniMiB-SHAR \cite{UniMiB-SHAR} (3 seconds records), via the extraction algorithms, is presented in Table~\ref{tab:extracted_features_amount}.

\begin{table}[htbp]
    \caption{Number of extracted features}
    \vspace{0.3cm}
    \label{tab:extracted_features_amount}
    \centering
    \color{black}
\begin{tabular}{m{0.28\linewidth}
>{\centering\arraybackslash}m{0.05\linewidth}
>{\centering\arraybackslash}m{0.05\linewidth}
>{\centering\arraybackslash}m{0.14\linewidth}
>{\centering\arraybackslash}m{0.05\linewidth}
>{\centering\arraybackslash}m{0.14\linewidth}
>{\centering\arraybackslash}m{0.03\linewidth}
>{\centering\arraybackslash}m{0.05\linewidth}}
\toprule
\textbf{No. of Extracted Features} & \textbf{CWT} & \textbf{SVM} & \textbf{Total $\mid$SVM$\mid$} & \textbf{SMA} & \textbf{Data Range} & \textbf{SE} & \textbf{Total} \\
\midrule
\textbf{Gibson et al. \cite{Gib16}} & 303 & 101 & 1 & 1 & 3 & 3 & 412\\
\midrule
\textbf{UniMiB-SHAR \cite{UniMiB-SHAR}} & 453 & 151 & 1 & 1 & 3 & 3 & 612\\
\bottomrule
\end{tabular}
\end{table}

Feeding individual features into previously mentioned classifiers generates good accuracy, however, concatenating multiple features generate better results as shown in Table~\ref{tab:different_features_testing}.

\begin{table}[H]
    \caption{Testing different features combinations accuracies}
    \vspace{0.3cm}
    \label{tab:different_features_testing}
    \centering
    \begin{threeparttable}
\renewcommand{\arraystretch}{0.8}
\begin{tabular}{
>{\bfseries}m{0.53\linewidth}
>{\centering\arraybackslash}m{0.08\linewidth}
>{\centering\arraybackslash}m{0.08\linewidth}
>{\centering\arraybackslash}m{0.08\linewidth}
>{\centering\arraybackslash}m{0.08\linewidth}
}
\toprule
\multirow{2}{*}{\textbf{Feature Combination}} 
& \multicolumn{4}{c}{\textbf{Accuracy (\%)}} \\
\cline{2-5}
& KNN\tnote{b} & ENN\tnote{b} & BDT & VM \\ 
\midrule
Total $\mid$SVM$\mid$ & 76.81 & 75.94 & 76.52 & 76.23 \\
SMA & 81.45 & 80.29 & 78.55 & 80.87 \\
SVM & 82.32 & 86.67 & 76.23 & 86.38 \\
SMA \& SVM & 83.77 & 85.51 & 85.51 & 85.80 \\
Data Range & 85.51 & 84.35 & 81.74 & 84.64 \\
Raw Accelerometer Data & 86.96 & 91.88 & 88.70 & 92.75 \\
CWT\tnote{a} & 92.75 & 93.91 & 91.59 & 93.91 \\
SMA, SE \& SVM & 92.46 & 95.94 & 88.99 & 96.23 \\
CWT\tnote{a}\quad \& SVM & 94.20 & 95.65 & 89.28 & 95.07 \\
SE & 94.49 & 94.49 & 93.04 & 95.07 \\
CWT\tnote{a}\quad \& SE & 94.78 & 94.78 & 90.43 & 94.78 \\
CWT\tnote{a}\quad \& SMA & 96.23 & 97.10 & 90.72 & 96.81 \\
CWT\tnote{a}\quad \& Total $\mid$SVM$\mid$ & 95.94 & 97.10 & 93.33 & 96.52 \\
SMA, SE \& Data Range & 95.36 & 94.78 & 92.17 & 95.36 \\
CWT\tnote{a}\ , SVM, Data Range, SMA, SE \& Total $\mid$SVM$\mid$ & \multicolumn{1}{c}{96.52} & \multicolumn{1}{c}{96.81} & \multicolumn{1}{c}{91.88} & \multicolumn{1}{c}{96.81} \\
CWT\tnote{a}\ , SE, SMA \& Total $\mid$SVM$\mid$ & 97.39 & 97.68 & 92.75 & 97.68 \\
CWT\tnote{a}\ , SE, SMA \& SVM & 97.10 & 98.55 & 92.17 & 98.26 \\
\bottomrule
\end{tabular}
\begin{tablenotes}
    \item[a] CWT features were calculated using the Biorthogonal2.2 wavelet mother function with a scale of 250 by trial and error. CWT generates 101 features/axis. 
    \item[b] $K$ = 3 \& $N$ = 3 for KNN \& ENN respectively.
\end{tablenotes}
\end{threeparttable}
\end{table}

For the CWT features, other wavelet functions are yielding similar accuracy results, with same scale (250), such as Haar, symlet(1-3), Daubechies(1-3), Meyer, etc. Thus, they can be used interchangeably. However, reducing the scale to be less than 100 shows reduction in classifiers' performance.

\begin{figure}[!ht]
    \centering
    \includegraphics[trim={1.55in 1.6in 1in 1.7in},clip,width=0.78\linewidth]{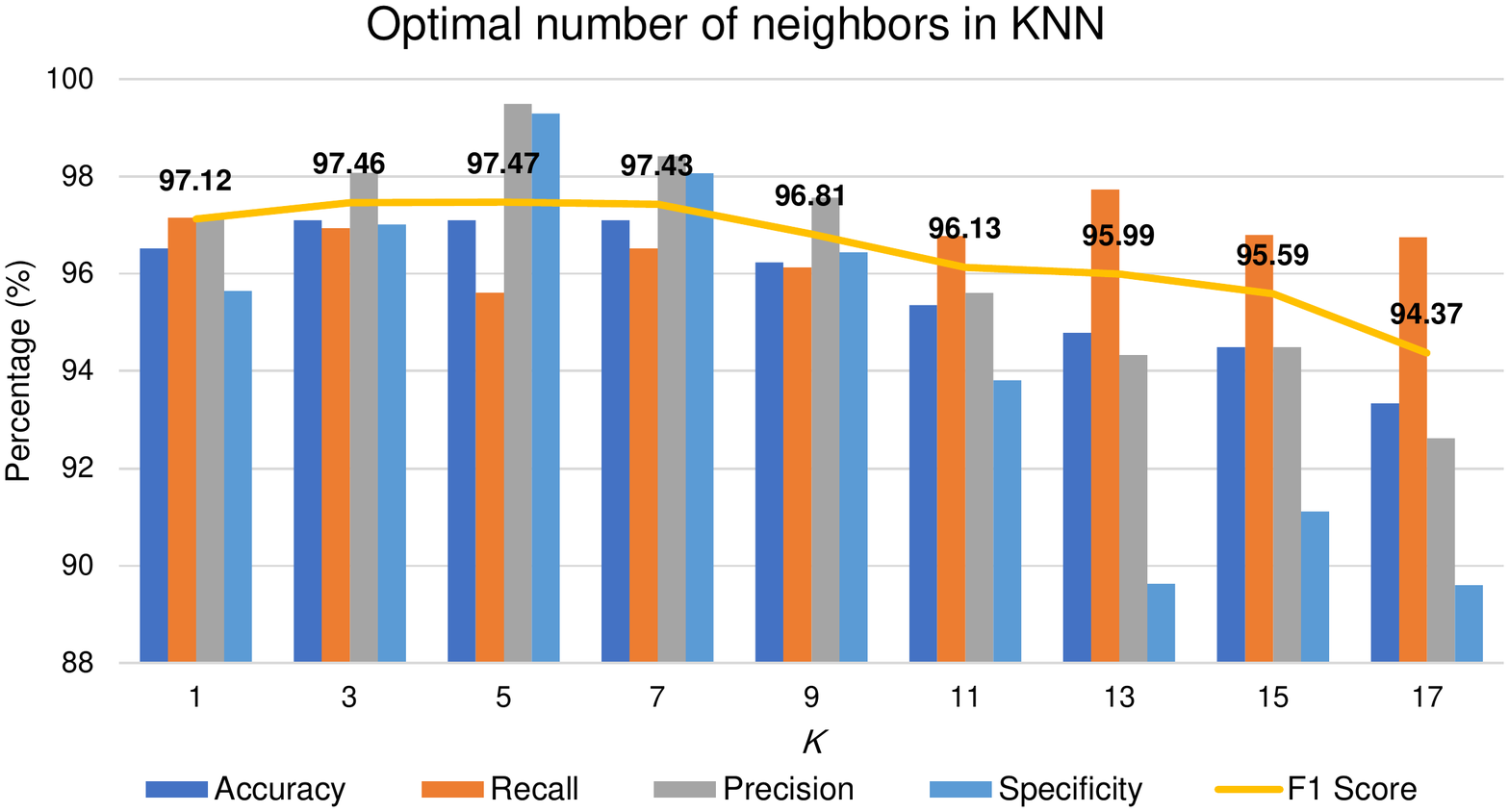}
    \caption{Testing KNN with different No. of neighbors}
    \label{fig:KNN_testing}
\end{figure}

\begin{figure}[!ht]
    \centering
    \includegraphics[trim={1.55in 1.6in 1in 1.7in},clip,width=0.78\linewidth]{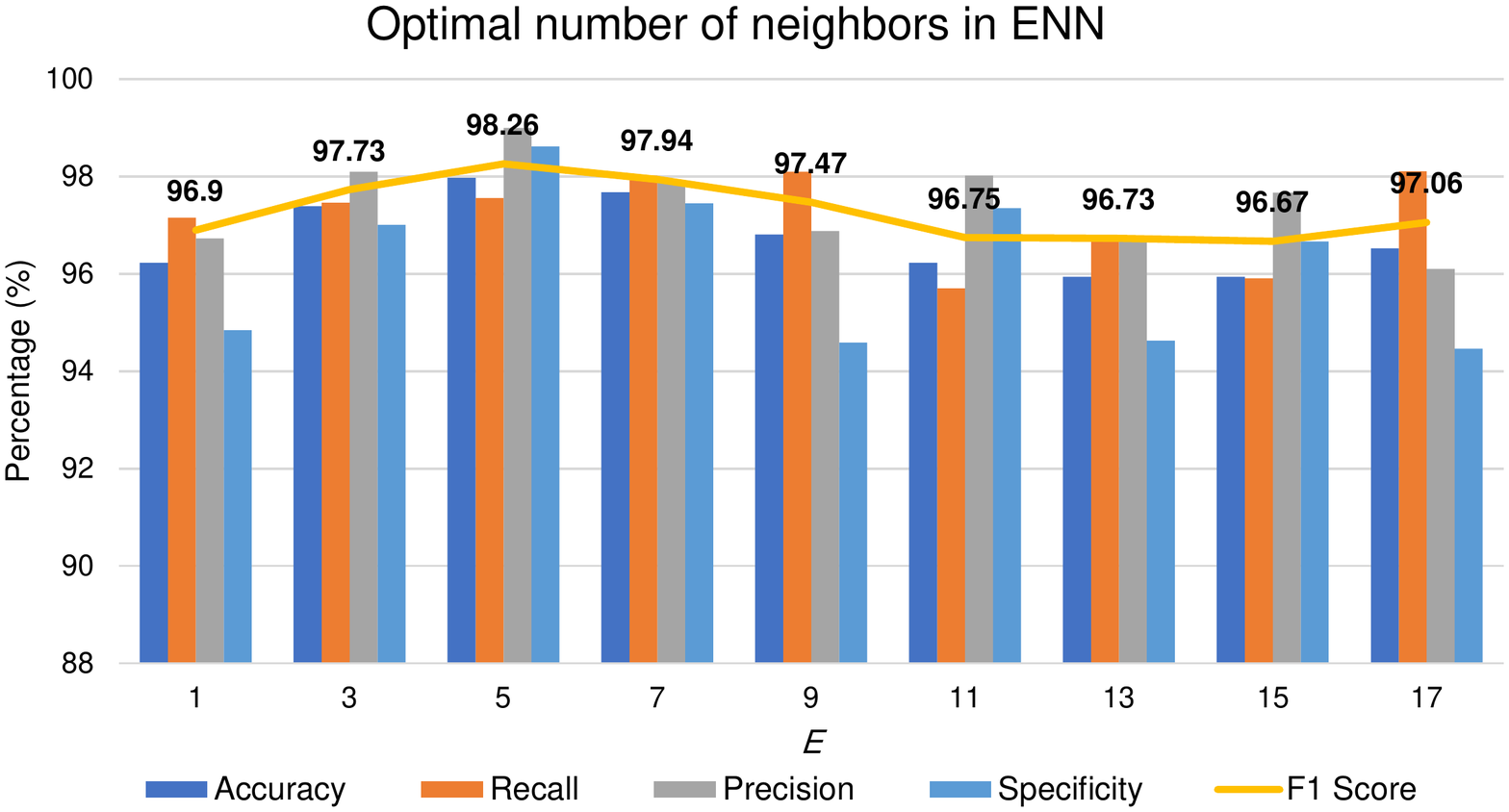}
    \caption{Testing ENN with different No. of neighbors}
    \label{fig:ENN_testing}
\end{figure}

The accuracy values depicted in Table~\ref{tab:different_features_testing} are ordered in an ascending order based on the classifiers’ overall performance. The last two rows show most promising results. Either one would be an excellent candidate to show next results, thus, the features in the last row are chosen. Feature normalization was examined for features in the last row but KNN and ENN performance has worsen. For BDT performance was relatively the same. Thus, feature normalization was not applied to the dataset. Choosing the correct number of neighbors considered for both KNN and ENN is a hyperparameter that significantly affects the aforementioned performance criteria and this can be clearly observed in Figure~\ref{fig:KNN_testing} and Figure~\ref{fig:ENN_testing}.




In Figure~\ref{fig:KNN_testing}, when $K \in$ \{3, 5, 7\}, the performance criteria show best results and the F\textsubscript{1} Score criterion affirms that as well. The same can be seen for ENN when $E \in$ \{3, 5, 7\} from Figure~\ref{fig:ENN_testing}. Using low number of neighbors for KNN and ENN is equivalent to overfitting on the training data. Increasing number of neighbors would reduce this overfitting making the classification algorithm generalize better when new records arrive. However, if the number of neighbors increases significantly, it would be equivalent to the underfitting scenario, meaning that the classifier is not trained well. Consequently, error on new incoming records will also be larger. Hence, this shows the need for tuning the number of neighbors to find the optimal one resulting in least classification error on the test set. The performance of ENN worsens in a slower pace than that of KNN, possibly showing the significance of ENN’s ``two-way communication''. 

The results in Figure~\ref{fig:KNN_testing} and Figure~\ref{fig:ENN_testing} indicate that the optimal number of neighbors is $K$ = 5 and $E$ = 5 for KNN and ENN, respectively. In addition, F\textsubscript{1} Score is calculated each iteration and the one presented in these figures is the mean of these five folds. To compare with the individual classifiers in \cite{Gib16}, KNN and ENN are achieving better results due to the deployed features concatenations where CWT is concatenated with SE, SMA and SVM. Although the number of features is higher than the one reported in \cite{Gib16}, the classification time is still adequate and is considered to be real-time as shown in Table~\ref{tab:performance_comparison}. Note that the recorded classification durations in Table~\ref{tab:performance_comparison} are acquired on a “Lenovo ideapad 500-15ISK” laptop with 8GB single-channel RAM, 1TB SSD, and an Intel Core i7-6500 dual-core processor with a clock speed of [2.50 – 2.60] GHz on MATLAB R2020a. 

\textcolor{black}{After analyzing the best feature vector combinations on the Gibson et al. \cite{Gib16} dataset, and to further validate our results, the larger dataset UniMiB-SHAR \cite{UniMiB-SHAR} has been used. The performance comparison on both datasets is shown in Table~\ref{tab:performance_comparison}, highlighting the classifiers' performance, with more variance on train/test ratio on the UniMiB-SHAR dataset due to its large size. It is worth noting that the reported results are the average over the results obtained from each fold. Moreover, the VM classification time is the summation of all three classifiers along the time it takes to compute the majority vote.}

\begin{table}[!ht]
    \caption{\textcolor{black}{Classifiers' performance criteria comparison}}
    \vspace{0.3cm}
    \label{tab:performance_comparison}
    \centering
    \begin{tabular}{
>{\bfseries}m{0.43\linewidth}
>{\centering\arraybackslash}m{0.1\linewidth}
>{\centering\arraybackslash}m{0.1\linewidth}
>{\centering\arraybackslash}m{0.1\linewidth}
>{\centering\arraybackslash}m{0.1\linewidth}
}
\toprule
\textbf{Performance Criteria} & \textbf{KNN} & \textbf{ENN} & \textbf{BDT} & \textbf{VM} \\ 
\midrule\midrule
\multicolumn{5}{c}{\textbf{Gibson et al. dataset \cite{Gib16}, 5 Folds, (70/30) Train/Test Ratio}} \\
\midrule

\textbf{Accuracy (\%)}                                        & 96.23 & 97.68 & 94.20 & 97.68 \\
\textbf{Recall (\%)}                                          & 96.72 & 98.63 & 95.14 & 98.63 \\
\textbf{Precision (\%)}                                       & 97.16 & 97.67 & 95.19 & 97.67 \\
\textbf{F\textsubscript{1} Score (\%)}                        & 96.94 & 98.13 & 95.15 & 98.13 \\
\textbf{Specificity (\%)}                                     & 95.45 & 96.25 & 92.59 & 96.25 \\
\textbf{Average Classification Time for a Single Record (ms)} & 5.78  & 0.36  & 0.31  & 6.50 \\
\bottomrule
\multicolumn{5}{c}{\textbf{UniMiB-SHAR Dataset \cite{UniMiB-SHAR}, 5 Folds, (70/30) Train/Test Ratio}} \\
\midrule
\textbf{Accuracy (\%)}                                         & 98.94 & 99.07 & 96.79 & 99.14 \\
\textbf{Recall (\%)}                                           & 98.84 & 98.98 & 95.53 & 99.09 \\
\textbf{Precision (\%)}                                        & 98.19 & 98.42 & 95.45 & 98.52 \\
\textbf{F\textsubscript{1} Score (\%)}                         & 98.51 & 98.70 & 95.49 & 98.80 \\
\textbf{Specificity (\%)}                                      & 98.99 & 99.12 & 97.49 & 99.17 \\
\textbf{Average Classification Time for a Single Record (ms)}  & 43.56 & 33.81 & 0.52 & 77.96 \\
\bottomrule
\multicolumn{5}{c}{\textbf{UniMiB-SHAR Dataset \cite{UniMiB-SHAR}, 10 Folds, (90/10) Train/Test Ratio}} \\
\midrule
\textbf{Accuracy (\%)}                                         & 99.09 & 99.15 & 96.70 & 99.19 \\
\textbf{Recall (\%)}                                           & 99.01 & 99.05 & 95.54 & 99.12 \\
\textbf{Precision (\%)}                                        & 98.47 & 98.58 & 95.21 & 98.63 \\
\textbf{F\textsubscript{1} Score (\%)}                         & 98.73 & 98.81 & 95.37 & 98.87 \\
\textbf{Specificity (\%)}                                      & 99.14 & 99.21 & 97.34 & 99.23 \\
\textbf{Average Classification Time for a Single Record (ms)}  & 54.53 & 43.31 & 0.54  & 98.45 \\

\bottomrule
\end{tabular}
\end{table}

If one of the classifiers is to be chosen and others to be discarded, a check on the overall performance for each classifier must be done. Table~\ref{tab:performance_comparison} shows all the performance results. \textcolor{black}{The time needed for extracting the utilized features (CWT, SVM, SMA, and SE) is in the [7.5ms - 9ms] range, which is highly dependent over the records' time length $t$ and the used sampling frequency $f_s$. In other words, it is dependent on the total number of samples within each record ($N = t{\times}f_s$). Nonetheless, the portrayed computational time values are very efficient from the real-time factor perspective. If $f_s$ is 50Hz for example, as it is the case with both utilized datasets, the feature extraction time for a single record is faster than acquiring a single triaxial acceleration sample from the record ($T_s$ = 20ms). To highlight the features vector's length, a total of 408 features for the Gibson et al. dataset are generated, while 608 features are generated for the UniMiB-SHAR dataset (refer back to Table~\ref{tab:feature_extraction_summary}).}

From observing Table~\ref{tab:performance_comparison}, \textcolor{black}{ENN is outperforming KNN on both datasets in the performance criteria especially on the smaller dataset size, due to its ability to significantly extract meaningful information from the available records. Moreover, it consumes less time during classification noting that the ENN has a pre-processing (training) phase that is done only once, while the KNN does not have any. From time-efficiency point-of-view, BDT is significantly outperforming both ENN and KNN as it builds a model that does not expand with higher number of records, but gets finely-tuned. Contrastingly, both ENN and KNN are model-free in the sense that the distance between the test record and the other training records need to be computed to predict the group it belongs to.}

\textcolor{black}{The huge number of provided records within the UniMiB-SHAR dataset allows for enhancing the classifiers' performance, as the classifiers have ``seen'' more variant records for ADL and falling scenarios. Thus, more tests on the performance of the aforementioned classifiers are created. It is also worth mentioning that ENN has a pre-processing phase that is heavily dependent on the number of records, and of high computational complexity $O(M^2log(M))$ to build the weighted KNN maps \cite{Tan15}, where $M$ is the number of records. It can be discarded as a training phase that will not be accounted into the classification time but it is mentioned here as following: i) 0.020 seconds for a training set of 159 records; ii) 284.5 seconds for a training set of 8,239 (70\% of 11,771) records; and iii) 470 seconds for a training set of 10,593 (90\% of 11,771) records.}

\textcolor{black}{Naturally, classification time for both classifiers significantly increases with the number of used records, while BDT remains relatively low because it is being trained based on the number of features within a single record, which is 408 for the case of Gibson et al. dataset and 608 in UniMiB-SHAR dataset's case.}

\subsection{\textcolor{black}{Comparison with the State-of-the-Art}}

\textcolor{black}{In Table~\ref{tab:market_comparison}, some of the commercial products available on the market are mentioned, where they offer an automatic fall detection feature. Similar products to the ones mentioned in Table~\ref{tab:market_comparison}, but not limited to, are \textit{LifeFone}, \textit{Bay Alarm Medical}, and \textit{GreatCall} \cite{ConsumerAffairs}. It is worth noting that these systems do have automatic fall detection scheme that either is an add-on feature mentioned explicitly, or embedded within the monthly fee that they collect. Moreover, the majority of these products also offer a 24/7 call service for the elderlies to talk with, and they come with a push button to request help in times of distress. MyNotifi is different than the other products in three ways: i) Although they provide automatic fall detection and communication with relatives, they do not have 24/7 call service, which can be thought of to be autonomous; ii) The system is much cheaper than others, as it is only a single-time payment while the others are monthly-based as long as the elderly is using the service; and iii) most importantly, they give details on the used algorithms, ANN, and obtained system accuracy, 96.2\%. Ours would be a single payment system, where if Shimmer3ECG is utilized as the sensory device, the system would cost around \$604, and it would cost \$59 if a smartphone is used as the sensor (\$59 for the ODROID-XU4), assuming that every individual already has a smartphone. The reason we utilized Shimmer3ECG in the first place is to include ECG data, which can work as a complementary source of information to the accelerometer data. It would prove its vitalness if no fall has occurred but the elderly is in a critical condition due to a health related-issue. However, if ECG information is not needed in a certain installment, it would be possible to replace it with the elderly's smartphone to collect the accelerometer data, similar to the UniMiB-SHAR dataset.}

\begin{table}[!ht]
    \caption{\textcolor{black}{Comparison with systems available in markets}}
    \vspace{0.3cm}
    \label{tab:market_comparison}
    \centering
    \begin{tabular}{
>{\bfseries}m{0.25\linewidth}
>{\centering\arraybackslash}m{0.25\linewidth}
>{\centering\arraybackslash}m{0.25\linewidth}
>{\centering\arraybackslash}m{0.15\linewidth}
}
\toprule
\textbf{Brand} & \textbf{Fall Detection} & \textbf{Placement} & \textbf{Total Price}\\ 
\midrule
\textbf{Mini Guardian \cite{MedicalGuardian}}         & Yes (\$10 monthly) & Pendant/Pocket (Other models exist) & +\$750 (yearly)  \\

\textbf{On The Go \cite{MedicalAlert}} & Yes (\$10 monthly) & Pendant/Pocket  & +\$430 (yearly)\\

\textbf{HomeSafe with AutoAlert/ GoSafe 2 \cite{PhilipsLifeline}} & Yes & Pendant/Pocket  & +\$540 (yearly)\\

\textbf{MobileHelp Solo \cite{MobileHelp}} & Yes & Pendant/Pocket & $\sim$\$395 (yearly)\\

\textbf{MyNotifi \cite{MyNotifi}} & Yes (96.2\% Accuracy) & Hip/Wrist  & \$200\\

\textbf{Ours (Sensor: Shimmer3ECG Unit)} & Yes & Hip  & \$604 \\

\textbf{Ours (Sensor: Smartphone)} & Yes & Pocket & \$59 \\

\bottomrule
\end{tabular}
\end{table}

\textcolor{black}{The reason behind mentioning these systems is to emphasize that fall detection issue is an actual and serious concern that many companies are tackling, however, can be quite expensive, especially for financially-incapable senior individuals.}

\textcolor{black}{Moreover, in Table~\ref{tab:state_of_the_art_comparison}, a comparison with the state-of-the-art is made, specifically in terms of the results that the algorithm portrays, and the ML algorithm it uses over the UniMiB-SHAR dataset. Firstly, the authors who published the dataset set a high bar for other researchers to exceed by obtaining an accuracy of 98.71\% for the AF-2 classification task \cite{UniMiB-SHAR}. Shahiduzzaman et al. \cite{Shahiduzzaman2019} used the UniMiB-SHAR dataset to create longer streams of records and were combined with camera-imagery input into an SVM\_ML algorithm. On the other hand, Ivascu et al. \cite{Ivascu2017} and Casilari et al. \cite{Casilari_2020} both used an ANN-based model achieving an accuracy of 96.73\% and 91.09\%, respectively. Delgado-Escaño et al. \cite{Ruben2020} utilized KNN ML algorithm where the results exceeded the ones mentioned earlier. Ours is simulated over the UniMiB-SHAR dataset where the results portray ENN as the best classifier over the 90/10 train/test ratio averaged over the 10-fold cross-validation, taken from Table~\ref{tab:performance_comparison}. As can be observed from the table, the results shown by ENN, along with the engineered features, are surpassing the state-of-the-art in the fall and non-fall classification task (AF-2), highlighting the novelty of our work.}

\begin{table}[H]
    \caption{\textcolor{black}{Comparison with the state-of-the-art over UniMiB-SHAR dataset (AF-2 Task)}}
    \vspace{0.3cm}
    \label{tab:state_of_the_art_comparison}
    \centering
\begin{tabular}{
>{\bfseries}m{0.2\linewidth}
>{\centering\arraybackslash}m{0.11\linewidth}
>{\centering\arraybackslash}m{0.06\linewidth}
>{\centering\arraybackslash}m{0.06\linewidth}
>{\centering\arraybackslash}m{0.06\linewidth}
>{\centering\arraybackslash}m{0.06\linewidth}
>{\centering\arraybackslash}m{0.06\linewidth}
>{\centering\arraybackslash}m{0.18\linewidth}
}
\toprule
\multirow{2}{*}{\textbf{Study}} & \multirow{2}{*}{\textbf{Algorithm}} & \multicolumn{5}{c}{\textbf{Metrics}} & \multirow{2}{*}{\textbf{Data Type}} \\ 
& & \rotatebox{0}{Ac (\%)} & \rotatebox{0}{Re (\%)} & \rotatebox{0}{Pr (\%)} & \rotatebox{0}{F\textsubscript{1} (\%)} & \rotatebox{0}{Sp (\%)} &  \\ 
\midrule

\textbf{Micucci et al. \cite{UniMiB-SHAR}} & SVM\_ML & 98.71 & --- & --- & --- & --- & Accelerometer\\
\midrule

\textbf{Shahiduzzaman et al. \cite{Shahiduzzaman2019}} & SVM\_ML & 96.67 & 96.67 & 98.38 & 97.52 & --- & Accelerometer + Camera\\
\midrule

\textbf{Ivascu et al. \cite{Ivascu2017}} & DNN & 96.73 & --- & --- & --- & ---  & Accelerometer\\
\midrule

\textbf{Casilari et al. \cite{Casilari_2020}} & CNN & 91.09 & 71.71 & --- & --- & 97.53 & Accelerometer\\
\midrule

\textbf{Delgado-Escaño et al. \cite{Ruben2020}} & KNN & 97.08 & --- & --- & --- & --- & Accelerometer\\
\midrule

\textbf{Ours} & \textbf{ENN} & \textbf{99.15} & \textbf{99.05} & \textbf{98.58} & \textbf{98.81} & \textbf{99.21} & \textbf{Accelerometer}\\

\bottomrule
\end{tabular}
\end{table}

\textcolor{black}{It is worth noting that WSD or smartphone-based systems are advantageous from multiple aspects when compared to other available systems. For instance, from privacy point-of-view, they are more private when compared to camera-based systems, where the elderlies would feel their privacy being violated. In comparison to other systems such as floor sensors, radars, WiFi-CSI, privacy can be considered maintained, however, the elderly can still be located, by an eavesdropper, within the household since the acquired data are spatially correlated. From a financial and logistics point-of-view, WSD-based systems are easier to install as all other systems require sensors that are either expensive, e.g. multiple cameras in each room, or restrictive, e.g. cameras, floor sensors, radars, and WiFi-CSI-based systems that require elderlies to be inside a household. WSD and smartphone-based systems are much less restrictive when it comes to mobility, as elderlies can carry them wherever they go. This allows them to leave the house more frequently, and have more physical activity in more refreshing environments instead of being imprisoned 24/7 within their households. From an accuracy point-of-view, all other systems are more prone to predicting false positives as they are more susceptible to noise coming from within the monitored environment. For instance, a relative or a pet can induce artifacts to the environment where the classifier might consider a mixture of these signals as a fall, simply because the environment is noisy, putting the elderly's life in significant risk. On contrast, a fall might not be detected (false negative) due to the noise within the environment, which can exhaust caregivers or hospitals if happens frequently. This is not the case with WSD and smartphone-based systems as they collect the information related to the elderly only, disregarding potential sources of noise within the environment. However, two main disadvantages that come to the surface when using WSD and smartphone-based systems are: 1) they can easily become uncomfortable; and 2) their reliance on battery charge. Lastly, within WSD-based systems, some of them use other sensors, along with 3-axis accelerometer, due to their availability such as gyroscope and magnetometer as in \cite{Pie15}. Although the achieved results following our method are already superior, adding more sensory data can be extremely useful, which could further perfect the system performance and make it more robust against noise or bias coming from a single sensor.}

\section{Conclusion}\label{sec6}
To conclude, \textcolor{black}{in order to provide elderlies with 24/7 healthcare service to support them in the event of a fall,} multiple feature extraction and classification algorithms for the presented fall detection system were evaluated. The most promising classifiers were KNN and ENN, but ENN outperformed KNN in both the performance criteria and its processing time on both examined datasets. Moreover, both performed extremely well on the UniMiB-SHAR dataset, showing state-of-the-art results. BDT’s results are also good but further inspection on how to improve them is a must, as detecting falls should have extremely low error rate. VM also showed promising results but its output was dictated by the results of both KNN and ENN since both are showing relatively similar behavior. The usage of F\textsubscript{1} Score for classifiers’ performance comparison was also shown where the models with the highest F\textsubscript{1} Score showed the most promising results. \textcolor{black}{From the presented results, it is evident that the extracted features, centered around the CWT, along with the selected classifiers are performing extremely well in detecting falls (Re~=~+99\%). Challenges that can face WSD-based systems is regarding their uncomfortability, and the limited battery charge. For the former issue, manufacturers are tackling it by making WSDs that can be worn comfortably on the wrist as a watch, or around the neck as a pendant (in the case of a smartphone the issue is almost non-existing, as the smartphone can be placed inside the pocket). The latter issue is much more significant as the elderly is susceptible to falling when the device is charging. A potential solution would be to design the sensors to have long battery lives, and use multiple ones, so that when one is charging, the other can be used, preventing the elderly from being exposed to undetected falls.}

\section*{Acknowledgment}
This paper was made possible by the National Priorities Research Program (NPRP) grant No. 9-114-2-055 from the Qatar National Research Fund (a member of Qatar Foundation). In addition, the work of Al-Kababji is supported by the Qatar National Research Fund Graduate Sponsorship Research Award (GSRA6-2-0521-19034). The statements made herein are solely the responsibility of the authors.

\bibliography{References}

\begin{thebibliography}{10}
\expandafter\ifx\csname url\endcsname\relax
  \def\url#1{\texttt{#1}}\fi
\expandafter\ifx\csname urlprefix\endcsname\relax\def\urlprefix{URL }\fi
\expandafter\ifx\csname href\endcsname\relax
  \def\href#1#2{#2} \def\path#1{#1}\fi

\bibitem{Uni}
\href{https://population.un.org/wpp/Publications/Files/WPP2019_Highlights.pdf}{{World
  Population Prospects 2019: Highlights}} (Jun. 2019).
\newline\urlprefix\url{https://population.un.org/wpp/Publications/Files/WPP2019_Highlights.pdf}

\bibitem{Com16}
\href{https://www.comfortkeepers.com/home/info-center/senior-independent-living/seniors-and-falls-statistics-and-prevention}{{Seniors
  and Falls: Statistics and Prevention}} (May 2016).
\newline\urlprefix\url{https://www.comfortkeepers.com/home/info-center/senior-independent-living/seniors-and-falls-statistics-and-prevention}

\bibitem{COVID_19_CDC}
\href{https://www.cdc.gov/nchs/nvss/vsrr/covid{\_}weekly/index.htm{\#}AgeAndSex}{{COVID-19
  Provisional Counts - Weekly Updates by Select Demographic and Geographic
  Characteristics}} (2021).
\newline\urlprefix\url{https://www.cdc.gov/nchs/nvss/vsrr/covid{\_}weekly/index.htm{\#}AgeAndSex}

\bibitem{yamada2020effect}
M.~Yamada, Y.~Kimura, D.~Ishiyama, Y.~Otobe, M.~Suzuki, S.~Koyama, T.~Kikuchi,
  H.~Kusumi, H.~Arai, {Effect of the COVID-19 epidemic on physical activity in
  community-dwelling older adults in Japan: A cross-sectional online survey},
  The journal of nutrition, health \& aging 24~(9) (2020) 948--950.

\bibitem{de2020falls}
M.~{\'A}. De~La~C{\'a}mara, A.~Jim{\'e}nez-Fuente, A.~I. Pardos, {Falls in
  older adults: the new pandemic in the post COVID-19 era?}, Medical Hypotheses
  (2020).

\bibitem{goethals2020impact}
L.~Goethals, N.~Barth, J.~Guyot, D.~Hupin, T.~Celarier, B.~Bongue, {Impact of
  Home Quarantine on Physical Activity Among Older Adults Living at Home During
  the COVID-19 Pandemic: Qualitative Interview Study}, JMIR aging 3~(1) (2020)
  e19007.

\bibitem{UniMiB-SHAR}
D.~Micucci, M.~Mobilio, P.~Napoletano,
  \href{http://dx.doi.org/10.3390/app7101101}{{UniMiB SHAR: A Dataset for Human
  Activity Recognition Using Acceleration Data from Smartphones}}, Applied
  Sciences 7~(10) (2017) 1101.
\newblock \href {https://doi.org/10.3390/app7101101}
  {\path{doi:10.3390/app7101101}}.
\newline\urlprefix\url{http://dx.doi.org/10.3390/app7101101}

\bibitem{cao2012falld}
Y.~Cao, Y.~Yang, W.~Liu, {E-FallD: A fall detection system using android-based
  smartphone}, in: 2012 9th International Conference on Fuzzy Systems and
  Knowledge Discovery, IEEE, 2012, pp. 1509--1513.

\bibitem{alkababji19}
A.~{Al-Kababji}, L.~{Shidqi}, I.~{Boukhennoufa}, A.~{Amira}, F.~{Bensaali},
  M.~S. {Gastli}, A.~{Jarouf}, W.~{Aboueata}, A.~{Abdalla}, {IoT-Based Fall and
  ECG Monitoring System: Wireless Communication System Based Firebase Realtime
  Database}, in: 2019 IEEE SmartWorld, Ubiquitous \& Intelligence Computing,
  Advanced \& Trusted Computing, Scalable Computing \& Communications, Cloud \&
  Big Data Computing, Internet of People and Smart City Innovation
  (SmartWorld/SCALCOM/UIC/ATC/CBDCom/IOP/SCI), 2019, pp. 1480--1485.

\bibitem{Mia16}
S.~. {Miaou}, {Pei-Hsu Sung}, {Chia-Yuan Huang}, {A Customized Human Fall
  Detection System Using Omni-Camera Images and Personal Information}, in: 1st
  Transdisciplinary Conference on Distributed Diagnosis and Home Healthcare,
  2006. D2H2., 2006, pp. 39--42.

\bibitem{Bia15}
Z.-P. Bian, J.~Hou, L.-P. Chau, N.~Magnenat-Thalmann, {Fall Detection Based on
  Body Part Tracking Using a Depth Camera}, IEEE Journal of Biomedical and
  Health Informatics 19 (2015) 430--439.

\bibitem{Sto15}
E.~E. Stone, M.~Skubic, {Fall Detection in Homes of Older Adults Using the
  Microsoft Kinect}, IEEE Journal of Biomedical and Health Informatics 19
  (2015) 290--301.

\bibitem{panahi2018human}
L.~Panahi, V.~Ghods, {Human fall detection using machine vision techniques on
  RGB--D images}, Biomedical Signal Processing and Control 44 (2018) 146--153.

\bibitem{Abo18}
A.~Abobakr, M.~Hossny, S.~Nahavandi, {A Skeleton-Free Fall Detection System
  From Depth Images Using Random Decision Forest}, IEEE Systems Journal 12
  (2018) 2994--3005.

\bibitem{mazurek2018use}
P.~Mazurek, J.~Wagner, R.~Z. Morawski, {Use of kinematic and
  mel-cepstrum-related features for fall detection based on data from infrared
  depth sensors}, Biomedical Signal Processing and Control 40 (2018) 102--110.

\bibitem{Zig09}
Y.~Zigel, D.~Litvak, I.~Gannot, {A Method for Automatic Fall Detection of
  Elderly People Using Floor Vibrations and Sound—Proof of Concept on Human
  Mimicking Doll Falls}, IEEE Transactions on Biomedical Engineering 26 (2009)
  2858--2867.

\bibitem{Dah17}
M.~Daher, A.~Diab, M.~E. B.~E. Najjar, M.~A. Khalil, F.~Charpillet, {Elder
  Tracking and Fall Detection System Using Smart Tiles}, IEEE Sensors Journal
  17 (2017) 469--479.

\bibitem{Rim10}
H.~Rimminen, J.~Lindstrom, M.~Linnavuo, R.~Sepponen, {Detection of falls among
  the elderly by a floor sensor using the electric near field}, IEEE
  Transactions on Information Technology in Biomedicine 14 (2010) 1475--1476.

\bibitem{Pet12}
P.~{Karsmakers}, T.~{Croonenborghs}, M.~{Mercuri}, D.~{Schreurs}, P.~{Leroux},
  Automatic in-door fall detection based on microwave radar measurements, in:
  2012 9th European Radar Conference, 2012, pp. 202--205.

\bibitem{Gar15}
C.~Garripoli, M.~Mercuri, P.~Karsmakers, P.~J. Soh, G.~Crupi, G.~A.~E.
  Vandenbosch, C.~Pace, P.~Leroux, D.~Schreurs, {Embedded DSP-Based Telehealth
  Radar System for Remote In-Door Fall Detection}, IEEE Journal of Biomedical
  and Health Informatics 1 (2015) 19.

\bibitem{SuB15}
B.~Y. Su, K.~C. Ho, M.~J. Rantz, M.~Skubic, {Doppler Radar Fall Activity
  Detection Using the Wavelet Transform}, IEEE Transactions on Biomedical
  Engineering 62 (2015) 865--875.

\bibitem{Jok18}
B.~Jokanovic, M.~Amin, {Fall Detection Using Deep Learning in Range-Doppler
  Radars}, IEEE Transactions on Aerospace and Electronic Systems 54 (2018)
  180--189.

\bibitem{Wan17}
Y.~Wang, K.~Wu, L.~M. Ni, {WiFall: Device-Free Fall Detection by Wireless
  Networks}, IEEE Transactions on Mobile Computing 16 (2017) 581--594.

\bibitem{Wan171}
H.~Wang, D.~Zhang, Y.~Wang, J.~Ma, Y.~Wang, S.~Li, {RT-Fall: A Real-Time and
  Contactless Fall Detection System with Commodity WiFi Devices}, IEEE
  Transactions on Mobile Computing 16 (2017) 511--526.

\bibitem{Hua19}
M.~{Huang}, J.~{Liu}, Y.~{Gu}, Y.~{Zhang}, F.~{Ren}, X.~{Wang}, J.~{Li}, Your
  wifi knows you fall: A channel data-driven device-free fall sensing system,
  in: ICC 2019 - 2019 IEEE International Conference on Communications (ICC),
  2019, pp. 1--6.

\bibitem{Cip17}
E.~Cippitelli, F.~Fioranelli, E.~Gambi, S.~Spinsante, {Radar and RGB-Depth
  Sensors for Fall Detection: A Review}, IEEE Sensors Journal 17 (2017)
  3585--3604.

\bibitem{Lee15}
J.~K. Lee, S.~N. Robinovitch, E.~J. Park, {Inertial Sensing-Based Pre-Impact
  Detection of Falls Involving Near-Fall Scenarios}, IEEE Transactions on
  Neural Systems and Rehabilitation Engineering 23 (2015) 258--266.

\bibitem{Pie15}
P.~Pierleoni, A.~Belli, L.~Palma, M.~Pellegrini, L.~Pernini, S.~Valenti, {A
  High Reliability Wearable Device for Elderly Fall Detection}, IEEE Sensors
  Journal 15 (2015) 4544--4553.

\bibitem{Gib16}
R.~M. Gibson, A.~Amira, N.~Ramzan, P.~Casaseca-de-la Higuera, Z.~Pervez,
  {Multiple comparator classifier framework for accelerometer-based fall
  detection and diagnostic}, Applied Soft Computing 39 (2016) 94--103.

\bibitem{gibson2017matching}
R.~M. Gibson, A.~Amira, N.~Ramzan, P.~Casaseca-de-la Higuera, Z.~Pervez,
  {Matching pursuit-based compressive sensing in a wearable biomedical
  accelerometer fall diagnosis device}, Biomedical signal processing and
  control 33 (2017) 96--108.

\bibitem{Azi17}
O.~Aziz, M.~Musngi, E.~J. Park, G.~Mori, S.~N. Robinovitch, {A comparison of
  accuracy of fall detection algorithms (threshold-based vs. machine learning)
  using waist-mounted tri-axial accelerometer signals from a comprehensive set
  of falls and non-fall trials}, Medical \& Biological Engineering \& Computing
  55 (2017) 45--55.

\bibitem{Eju17}
A.~Ejupi, M.~Brodie, S.~R. Lord, J.~Annegarn, S.~J. Redmond, K.~Delbaere,
  {Wavelet-Based Sit-To-Stand Detection and Assessment of Fall Risk in Older
  People Using a Wearable Pendant Device}, IEEE Transactions on Biomedical
  Engineering 64 (2017) 1602--1607.

\bibitem{Wan172}
C.~Wang, S.~J. Redmond, W.~Lu, M.~Stevens, S.~R. Lord, N.~H. Lovell, {Selecting
  Power-Efficient Signal Features for a Low-Power Fall Detector}, IEEE
  Transactions on Biomedical Engineering 64 (2017) 2729--2736.

\bibitem{deQ18}
T.~de~Quadros, A.~E. Lazzaretti, F.~K. Schneider, {A Movement Decomposition and
  Machine Learning-Based Fall Detection System Using Wrist Wearable Device},
  IEEE Sensors Journal 18 (2018) 5082--5089.

\bibitem{Saa19}
W.~Saadeh, S.~A. Butt, M.~A.~B. Altaf, {A Patient-Specific Single Sensor
  IoT-Based Wearable Fall Prediction and Detection System}, IEEE Transactions
  on Neural Systems and Rehabilitation Engineering 27 (2019) 995--1003.

\bibitem{Ozc16}
K.~Ozcan, S.~Velipasalar, P.~K. Varshney, {Autonomous Fall Detection With
  Wearable Cameras by Using Relative Entropy Distance Measure}, IEEE
  Transactions on Human-Machine Systems 47 (2016) 31--39.

\bibitem{Wib13}
W.~{Wibisono}, D.~N. {Arifin}, B.~A. {Pratomo}, T.~{Ahmad}, R.~M. {Ijtihadie},
  {Falls Detection and Notification System Using Tri-axial Accelerometer and
  Gyroscope Sensors of a Smartphone}, in: 2013 Conference on Technologies and
  Applications of Artificial Intelligence, 2013, pp. 382--385.

\bibitem{Kau15}
L.-J. Kau, C.-S. Chen, {A Smart Phone-Based Pocket Fall Accident Detection,
  Positioning, and Rescue System}, IEEE Journal of Biomedical and Health
  Informatics 19 (2015) 44--56.

\bibitem{Ker15}
H.~Kerdegari, S.~Mokaram, K.~Samsudin, A.~R. Ramli, {A pervasive neural network
  based fall detection system on smart phone}, Journal of Ambient Intelligence
  and Smart Environments 7 (2015) 221--230.

\bibitem{Sha19}
A.~Shahzad, K.~Kim, {FallDroid: An Automated Smart-Phone-Based Fall Detection
  System Using Multiple Kernel Learning}, IEEE Transactions on Industrial
  Informatics 15 (2019) 45--44.

\bibitem{ODR}
ameriDroid, \href{https://ameridroid.com/products/odroid-xu4}{{ODROID-XU4}}
  (2015).
\newline\urlprefix\url{https://ameridroid.com/products/odroid-xu4}

\bibitem{Con18}
\href{https://www.weisang.com/en/documentation/timefreqspectrumalgorithmscwt_en/}{{Continuous
  Wavelet Transform (CWT)}} (Feb. 2018).
\newline\urlprefix\url{https://www.weisang.com/en/documentation/timefreqspectrumalgorithmscwt_en/}

\bibitem{Rio91}
O.~Rioul, M.~Vetterli, {Wavelets and signal processing}, IEEE Signal Processing
  Magazine 8 (1991) 14--38.

\bibitem{Abd18}
A.~S.~A. {Sukor}, A.~{Zakaria}, N.~A. {Rahim}, {Activity recognition using
  accelerometer sensor and machine learning classifiers}, in: 2018 IEEE 14th
  International Colloquium on Signal Processing Its Applications (CSPA), 2018,
  pp. 233--238.

\bibitem{Erd16}
B.~Erdaş, I.~Atasoy, K.~Açici, H.~Oǧul, {Integrating Features for
  Accelerometer-based Activity Recognition}, Procedia Computer Science 58
  (2016) 522--527.

\bibitem{Tan15}
B.~Tang, H.~He, {ENN: Extended Nearest Neighbor Method for Pattern Recognition
  [Research Frontier]}, IEEE Computational Intelligence Magazine 10 (2015)
  52--60.

\bibitem{Ist16}
\href{https://www.researchgate.net/post/Is_there_an_ideal_ratio_between_a_training_set_and_validation_set_Which_trade-off_would_you_suggest}{{Is
  there an ideal ratio between a training set and validation set? Which
  trade-off would you suggest?}} (Mar. 2016).
\newline\urlprefix\url{https://www.researchgate.net/post/Is_there_an_ideal_ratio_between_a_training_set_and_validation_set_Which_trade-off_would_you_suggest}

\bibitem{ConsumerAffairs}
S.~Pope,
  \href{https://www.consumeraffairs.com/medical-alert-systems/fall-detection.html}{{7
  Best Medical Alerts with Fall Detection | ConsumerAffairs}} (2021).
\newline\urlprefix\url{https://www.consumeraffairs.com/medical-alert-systems/fall-detection.html}

\bibitem{MedicalGuardian}
{Medical Guardian},
  \href{https://www.medicalguardian.com/order-wizard/127}{{Order Now | Mini
  Guardian}} (2021).
\newline\urlprefix\url{https://www.medicalguardian.com/order-wizard/127}

\bibitem{MedicalAlert}
{Medical Alert},
  \href{https://www.medicalalert.com/product/on-the-go-2-day/}{{On the Go 2-Day
  Medical Alert System | Medical Alert}} (2021).
\newline\urlprefix\url{https://www.medicalalert.com/product/on-the-go-2-day/}

\bibitem{PhilipsLifeline}
{Philips Lifeline},
  \href{https://www.lifeline.philips.com/medical-alert-systems/homesafe-autoalert.html}{{HomeSafe
  with AutoAlert | Philips Lifeline}} (2021).
\newline\urlprefix\url{https://www.lifeline.philips.com/medical-alert-systems/homesafe-autoalert.html}

\bibitem{MobileHelp}
MobileHelp,
  \href{https://www.mobilehelp.com/products/mobilehelp-solo}{{MobileHelp Solo}}
  (2021).
\newline\urlprefix\url{https://www.mobilehelp.com/products/mobilehelp-solo}

\bibitem{MyNotifi}
MyNotifi,
  \href{https://www.mynotifi.com/index.php?route=information/information{\&}information{\_}id=15}{{Product
  Accuracy - MyNotifi Medical Alert Devices {\&} Accessories}} (2021).
\newline\urlprefix\url{https://www.mynotifi.com/index.php?route=information/information{\&}information{\_}id=15}

\bibitem{Shahiduzzaman2019}
K.~M. {Shahiduzzaman}, X.~{Hei}, C.~{Guo}, W.~{Cheng}, {Enhancing Fall
  Detection for Elderly with Smart Helmet in a Cloud-Network-Edge
  Architecture}, in: 2019 IEEE International Conference on Consumer Electronics
  - Taiwan (ICCE-TW), 2019, pp. 1--2.
\newblock \href {https://doi.org/10.1109/ICCE-TW46550.2019.8991972}
  {\path{doi:10.1109/ICCE-TW46550.2019.8991972}}.

\bibitem{Ivascu2017}
T.~{Ivascu}, K.~{Cincar}, A.~{Dinis}, V.~{Negru}, Activities of daily living
  and falls recognition and classification from the wearable sensors data, in:
  2017 E-Health and Bioengineering Conference (EHB), 2017, pp. 627--630.
\newblock \href {https://doi.org/10.1109/EHB.2017.7995502}
  {\path{doi:10.1109/EHB.2017.7995502}}.

\bibitem{Casilari_2020}
E.~Casilari, R.~Lora-Rivera, F.~García-Lagos,
  \href{http://dx.doi.org/10.3390/s20051466}{{A Study on the Application of
  Convolutional Neural Networks to Fall Detection Evaluated with Multiple
  Public Datasets}}, Sensors 20~(5) (2020) 1466.
\newblock \href {https://doi.org/10.3390/s20051466}
  {\path{doi:10.3390/s20051466}}.
\newline\urlprefix\url{http://dx.doi.org/10.3390/s20051466}

\bibitem{Ruben2020}
R.~Delgado-Escaño, F.~M. Castro, J.~R. Cózar, M.~J. Marín-Jiménez, N.~Guil,
  E.~Casilari,
  \href{http://www.sciencedirect.com/science/article/pii/S0169260719311770}{{A
  cross-dataset deep learning-based classifier for people fall detection and
  identification}}, Computer Methods and Programs in Biomedicine 184 (2020)
  105265.
\newblock \href {https://doi.org/https://doi.org/10.1016/j.cmpb.2019.105265}
  {\path{doi:https://doi.org/10.1016/j.cmpb.2019.105265}}.
\newline\urlprefix\url{http://www.sciencedirect.com/science/article/pii/S0169260719311770}

\end{thebibliography}

\end{document}